\documentclass[twocolumn]{aastex6}
\usepackage{natbib}
\usepackage{color}
\usepackage{soul}
\usepackage{graphicx}
\begin{document}
\title{Young cluster Berkeley~59 : Properties, evolution and star formation}
\author{Neelam Panwar \altaffilmark{1}, A. K. Pandey \altaffilmark{2}, 
Manash R. Samal \altaffilmark{3}, Paolo Battinelli \altaffilmark{4}, 
K. Ogura \altaffilmark{5}, D. K. Ojha \altaffilmark{6}, 
W. P. Chen \altaffilmark{3}, H. P. Singh \altaffilmark{1}}
\altaffiltext{1}{Department of Physics \& Astrophysics, 
University of Delhi,
Delhi-110007, India}
\altaffiltext{2}{Aryabhatta Research Institute of Observational Sciences (ARIES), Nainital - 263129, India}
\altaffiltext{3}{Graduate Institute of Astronomy, 
National Central University 300, Jhongli City, Taoyuan County - 32001, Taiwan}
\altaffiltext{4}{INAF, Osservatorio Astronomico di Roma Viale del Parco Mellini 84, I-00136 Roma, Italy} 
\altaffiltext{5}{Kokugakuin University, Higashi, Shibuya-ku, Tokyo - 1508440, Japan}
\altaffiltext{6}{Tata Institute of Fundamental Research, Mumbai (Bombay) - 400 005, India}
\begin{abstract}
Berkeley~59 is a nearby ($\sim$1 kpc) young cluster associated with the Sh2-171 
H{\sc ii} region. We present deep optical observations of the central $\sim$ 2.5 $\times$ 2.5 pc$^2$ 
area of the cluster, obtained with the 3.58-m Telescopio Nazionale Galileo. The $V$/($V$-$I$) 
color-magnitude diagram manifests a clear pre-main-sequence (PMS) population down to $\sim$ 0.2 M$_\odot$. 
Using the near-infrared  and optical colors of the low-mass PMS members we derive a global extinction 
of A$_V$= 4 mag and a mean age of $\sim$ 1.8 Myr, respectively, for the cluster. We constructed the initial mass 
function and found that its global slopes in the mass ranges of  0.2 - 28 M$_\odot$ and 0.2 - 1.5 M$_\odot$ 
are -1.33 and -1.23, respectively, in good agreement with the Salpeter value  in the solar neighborhood. 
We looked for the radial variation of the mass function and found that the slope is flatter in the inner region than in the outer region, indicating mass segregation. The dynamical status of the cluster suggests that the mass segregation 
is likely primordial. The age distribution of the PMS sources reveals that the younger sources 
appear to concentrate close to the inner region compared to the outer region of the cluster, a phenomenon 
possibly linked to the time evolution of star-forming clouds is discussed. Within the observed area, 
we derive a total mass of $\sim$ 10$^3$ M$_\odot$ for the cluster. Comparing the properties of Berkeley~59 
with other young clusters, we suggest it resembles more to the Trapezium cluster.
\end{abstract}

\section{Introduction} \label{sec:intro}
Since most stars in the Galaxy are formed in clusters \citep[e.g.,][]{lada03},
the processes responsible for cluster formation are important to include 
any consideration of the mechanisms of star formation. Despite recent advances that have been made 
in this field, including observational and theoretical work focusing on the different stages of star formation, 
a number of mechanisms remain unclear. It is still debated whether star formation is a fast, 
dynamic process \citep{elm00,elm07,har07} or 
slow \citep{tan06} proceeding for at least several dynamical timescales. 
Similarly, since most of the massive stars form in clusters it is  still unclear
whether initial mass function (IMF) is universal \citep{bast10,off14} or it depends on the environments as observed in some extreme 
starburst galaxies and young massive clusters \citep{sto05,van10,lu13}.
Massive stars potentially influence their surroundings through their energetic radiation and stellar winds by regulating the dynamics, density and temperature distribution of the gas \citep[e.g.,][]{koe08,ojha11,deh12,sam14,pan14,jose16}. 
Hence these could be dominating factors in shaping the fundamental
properties of the cluster such as IMF and 
total star formation efficiency. Also the evolution of circumstellar
disks around the young stars can be affected \citep[e.g.][and refernces therein]{dib10}. 

Observations of very young clusters, before they dynamically relaxed, can 
provide some indications of the initial conditions of their
formation. Therefore, a careful characterization of the basic properties of 
clustered systems at their early evolutionary stages is essential to 
put constraints on the formation of such systems.
  
Berkeley~59 (Be~59) is a young cluster \citep[$\sim$ 2 Myr;][]{pan08} located at the centre of the Cepheus OB4 stellar association and is surrounded
by the H{\sc ii} region Sh2-171 \citep{yan92}. It is a relatively nearby cluster (distance $\sim$ 1 kpc) showing variable reddening 
 with $E$($B$ - $V$) ranging from 1.4 to 1.8 mag \citep{maj08,pan08}.
It contains several massive stars (spectral type from O7 to B5) \citep{kun08,maj08,skif14,mai16}
which are still associated with the natal molecular clouds making it one of the potential nearby massive clusters. 
\citet{yan92} 
observed two dense molecular clumps (`C1' and `C2') at the western side of the Be~59 using
the J = 1 - 0 lines of $^{12}$CO and $^{13}$CO emission. 
They suggested that the dense gas is in contact with
the H{\sc ii} region Sh2-171, and the ionization front (IF) is driving shocks into the clumps and a new generation of stars may be triggered to form from the compressed gas layer. \citet{ros13} studied the region using WISE 
data and identified a small group of embedded stars (RA: 00h 00m 46s, DEC: +67$^\circ$32$^\prime$59$\arcsec$, J2000.0) which resides in a cloud within the clump `C2'. On the basis of velocity studies \citet{yan92} found that the local standard-of-rest 
velocity (V$_{LSR}$) of the clump `C2' 
(-16 km/s) matches that of Be~59 (-15.7 km/s) derived from observations of member stars \citep{liu89}. Hence, 
the clump appears to be physically related with Be~59.

Though the cluster has been investigated by several authors at
optical bands \citep{maj08,pan08,lat11,esw12}, most of these studies were limited to the high to intermediate mass
stars only. \citet{pan08} identified intermediate mass population of young stellar objects (YSOs) in the region 
using near-infrared 2MASS and slitless grism H$\alpha$ data. However, their optical photometry could detect stars upto V $\sim$ 18.5 mag corresponding to $\sim$ 1.5 M$_\odot$ at the adopted distance and extinction of the cluster. Recently, \citet{koe12,ros13,get17} identified candidate YSOs in the region, however, no adequate attention 
has been paid to the characterization
of low-mass young stellar population of the cluster by using deep optical photometric data.
Since the low-mass stars constitute the majority of the stellar population in the Galaxy, 
the low-mass stellar content and its IMF are essential to understand the nature of the star formation 
process and the properties of stellar systems. 

In this paper, we present new 
observations of the central area of Be~59 at optical 
wavelengths obtained with the 3.58-m Telescopio Nazionale Galileo (TNG). These new observations are the deepest to date for this
region and allow us to assess the stellar contents of the
cluster down to 0.2 M$_\odot$, and provide a better analysis of its properties and mass function, understanding on the dynamical status, and insight into its formation. 
In Section 2, we present the optical observations, the data reduction and the archive datasets utilized in this study. In Section 3, the identification 
of PMS population, extinction measurements, IMF and the mass-segregation and star-formation associated to the cluster are discussed. 
Finally, the results of the present work are summarized in Section 4.
\section{Observations and Data Reduction}
 Figure 1 shows the color-composite image (red: 4.6 $\micron$ from WISE, green: 2.2 $\micron$ from 2MASS, blue: 0.65 
$\micron$ from DSS2) of the Be~59 region.  
The boxes `A' and `B' ($\sim$ 8$^\prime$.6 $\times$ 
8$^\prime$.6 each) represent the area covered by our optical observations. `C1' and `C2' 
represent the locations of the peaks of the molecular clumps \citep[cf.][]{yan92}.
 
In Fig. 1, a clustering of stellar sources is clearly apparent at the center of the image. 
The figure also displays several red WISE sources, indicating the presence of a likely young stellar population, 
still deeply embedded in the molecular cloud. The diffuse optical emission is mainly seen at the center of the cluster indicating
low-extinction at the central direction compared to the outskirts where molecular cloud is present. In the present work, we used optical observations of `region A' to examine the cluster properties of Be~59, while observations of `region B' are used to understand the origin of star formation 
observed towards C2. 
\begin{figure*}
\centering
\includegraphics[width=17cm]{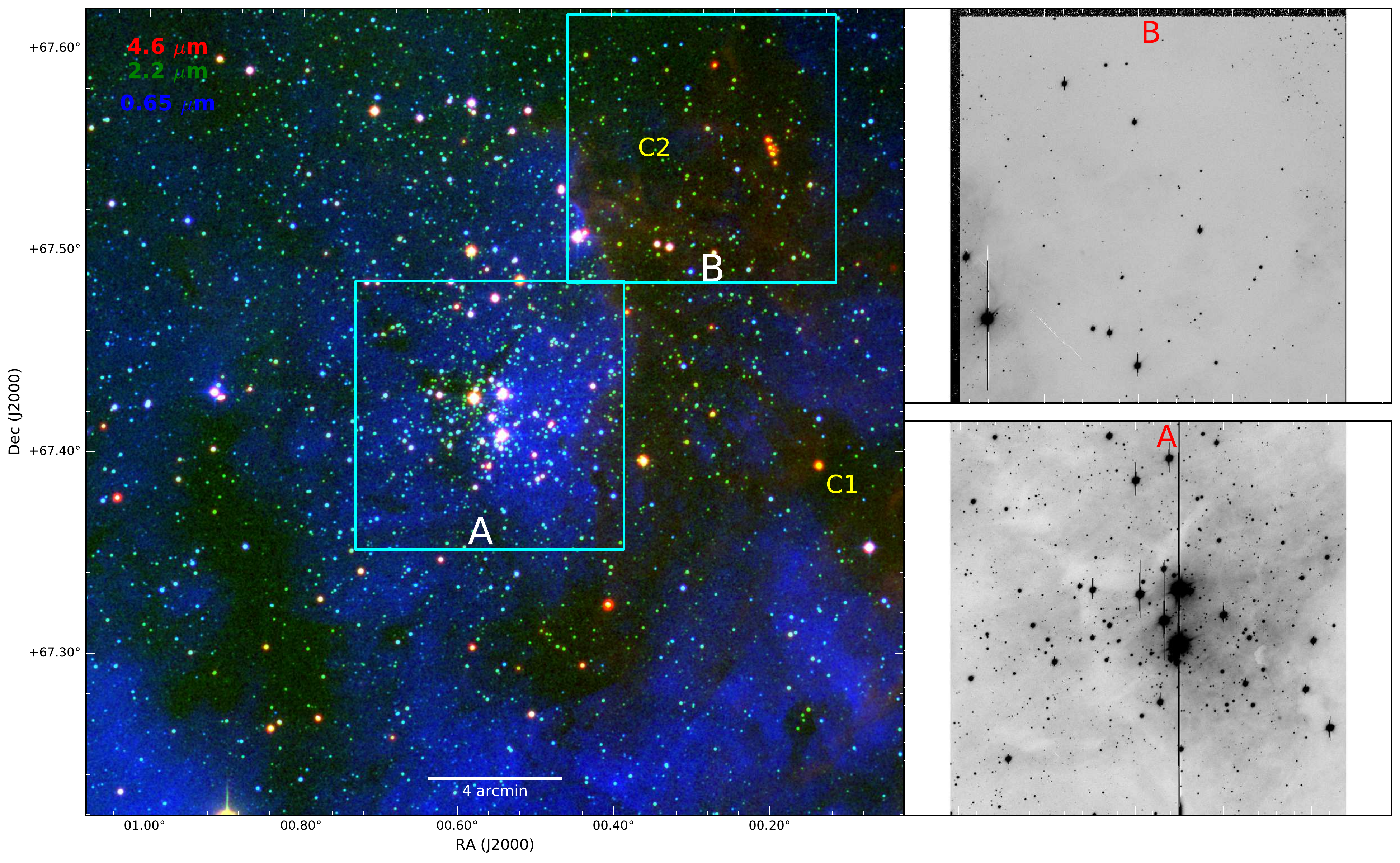}
\caption{Three-color image (red: 4.6 $\micron$, green: 2.2 $\micron$, blue: 0.65 $\micron$) of the 
Be~59 region. The two boxes (marked as `A' and `B')  
represent the DOLORES observations. 
`C1' and `C2' are the locations of the two dense clumps identified by \citet{yan92}. The optical images of the regions A and B are shown in grey scale also. 
}
\end{figure*}
\subsection{Optical Observations}

The {\it VI} observations of the two regions of the cluster Be~59 (see Fig. 1), region `A' (RA: 00h 02m 14.0s, DEC:+67$^\circ$25$^\prime$08$\arcsec$.0) 
and region `B' (RA: 00h 01m 08.0s, DEC:+67$^\circ$33$^\prime$03${\arcsec}$.0), 
and a control field having the same area (centered at RA: 00h 09m 48.0s, DEC: +68$^\circ$07$^\prime$58$\arcsec$.0) were obtained by 
using the Device Optimized for the LOw RESolution (DOLORES) instrument mounted on the TNG. 
DOLORES is a focal reducer instrument equipped with a 2048 $\times$ 2048 square 
pixel CCD detector, with a plate scale of 0.252$^{\prime\prime}$/pixel and a field of view (FoV) 
of 8.6$^\prime$ $\times$ 8.6$^{\prime}$. During the 
observations the average seeing was $\sim$ $1^{\prime\prime}$. The details of the observations are given in table 1.
\begin{table*}
\caption{Log of observations} 
\begin{tabular}{|p{.7in}|p{.6in}|p{2.0in}|p{1.2in}|p{0.8in}|}
\hline
$\alpha_{(2000)}$ & $\delta_{(2000)}$ & Filter \& Exposure(sec)$\times$no. of frames& Date of observations & Regions\\
(h:m:s)           & (d:m:s)           &  &(yr-mm-dd)              &\\
\hline
00:02:14.0 &+67:25:08.0      & V:275$\times$6;I:275$\times$6 &  2009-10-12 & Region `A'\\
00:02:14.0 &+67:25:08.0      & V:60$\times$3;I:60$\times$3 &  2010-08-31 & Region `A'\\
00:01:08.0 &+67:33:03.0      & V:275$\times$6,60$\times$3;I:275$\times$6,6s$\times$3 &  2010-08-31& Region `B'\\
00:09:48.0 &+68:07:58.0      & V:275$\times$6,60$\times$3;I:275$\times$6,60$\times$3 &  2010-08-31& Field region\\
\hline
\end{tabular}
\label{tab1}
\end{table*}
 
A number of bias and flat frames were also taken during the observations. 
The pre-processing of the data frames was done by using various tasks 
available under the $IRAF$ data reduction software package. The fringing patterns 
were present in the $I$ band frames which were removed by using the 
$rmfringe$ task available in $IRAF$. The photometric 
measurements of the stars were performed by using $DAOPHOT-II$ 
software package \citep{ste87}. The point spread function (PSF) was obtained for each 
frame by using several uncontaminated stars and the PSF photometry of all the sources was obtained using the ALLSTAR task in $IRAF$ \citep[e.g.,][]{ojha09,jos17}.

The instrumental magnitudes were converted to the standard values by using the secondary standards from 
\citet{lat12}. The photometric accuracies depend on the brightness of the stars. We finally consider only those sources with uncertainty $<$0.2 mag in $V$ and $I$. Our data (detection limit, $V$ $\sim$ 24 mag) are $\sim$ 5.5 mag deeper than that of \citet[][$V$ $\sim$ 18.5 mag]{pan08}.

\subsection{Data Completeness}
To derive cluster properties, it is necessary to take into account the 
incompleteness in the observed data that may occur for various reasons, e.g., crowding of the stars, diffuse emission, variable extinction etc. 
A quantitative evaluation of the completeness of the photometric data with respect to the brightness 
and the position on a given frame is necessary to convert the observed luminosity function (LF) 
to a true LF. We used the ADDSTAR routine of $DAOPHOT-II$ to determine the completeness factor (CF). 
The procedure has been outlined in detail in our earlier works \citep[e.g.,][]{pan01,cha11}. 
Briefly, we randomly added the artificial stars to both $V$ and $I$ images in such a way that they 
have similar geometrical locations. The luminosity distribution of artificial stars is chosen 
in such a way that more stars are inserted towards the fainter magnitude bins. The frames were 
reduced by using the same procedure used for the original frame. The ratio of the number of 
stars recovered to that added in each magnitude interval gives the CF as a function of magnitude. 
The minimum value of the CF of the pair (i.e., the $V$ and $I$ band observations), 
given in Table \ref{cf_opt}, is used to correct the data for incompleteness. The incompleteness of the data 
increases with increasing magnitude as expected. As can be seen from the table, in $V$-band  
the completeness of the cluster region is better than 80\% level at $\sim$ 23.5 mag. 
\begin{table}
\caption{Completeness factor of the photometric data in the cluster region and control field.}
\label{cf_opt}
\begin{tabular}{cccc} \hline
V range (mag)& cluster region& control field \\
\hline
17.0 - 17.5&1.00 & 1.00\\
17.5 - 18.0&1.00 & 1.00\\
18.0 - 18.5&1.00 & 1.00\\
18.5 - 19.0&1.00 & 1.00\\
19.0 - 19.5&0.99 & 0.99\\
19.5 - 20.0&0.98 & 0.98\\
20.0 - 20.5&0.97 & 0.97\\
20.5 - 21.0&0.95 & 0.96\\
21.0 - 21.5&0.93 & 0.96\\
21.5 - 22.0&0.92 & 0.95\\
22.0 - 22.5&0.90 & 0.92\\
22.5 - 23.0&0.87 & 0.90\\
23.0 - 23.5&0.83 & 0.87\\
23.5 - 24.0&0.81 & 0.84\\
\hline    
\end{tabular}
\end{table}
\subsection{YSOs from Infra-red and X-ray Observations}
During their early phase, stars possess cirmcumstellar disks and hence show the excess emission in the IR 
wavelengths. The IR excess property of YSOs can be used to identify and classify them in different evolutionary classes, i.e., Class~0, Class~I, 
Class~II and Class III \citep{and95}. 
Whereas Class~`II' sources are surrounded 
by optically thick disks and also exhibit near infrared (NIR) to far infrared (FIR) excess emission, Class~`III' sources have optically thin disks or
 they may be disk anemic, so possess little or no excess in IR \citep{lada87,and95,lada06,luh12r}. Hence, IR observations are less 
sensitive to identify Class~III sources. However, Class~II, Class~III sources (referred as PMS stars) are often strong X-ray emitters and very luminous in the X-ray regime compared to their main-sequence 
(MS) counterparts \citep[e.g.,][]{fei99,get05,pre05,gue07,getman12}. 
Hence, X-ray surveys of star forming regions are used to uncover YSO populations. 
Particularly, for the detection of disk anemic YSOs, i.e., Class~III sources, X-ray observations complement the 
YSO sources obtained from IR observations and a better census of the low-mass stellar content of clusters can be achieved  
\citep[e.g.,][]{pre05,fei13,bro13,pan13,pand14,pre17}.  
 Recently, \citet{get17} carried out the {\it Star Formation in Nearby Clouds (SFiNCs)} project 
to provide detailed studies of the young stellar populations and star cluster formation in nearby star forming regions (SFRs)
 including Be~59. They identified $\sim$ 700 young stars using 2MASS (JH$K_s$), 
$Spitzer$-IRAC (3.6 $\micron$, 4.5 $\micron$) and X-ray data 
from $Chandra$ space telescope, which were classified as sources with disks, without disks and probable members. Our optical observations are well within their area of study. 
Hence, we have used their YSO catalog in this work to guide our analysis and, in particular, to 
identify the PMS stars of the cluster as well as to characterize the cluster. While NIR, MIR and X-ray observations are suitable to identify
PMS stars, the characterization of these stars (e.g., age and mass determinations) can be 
done more accurately by optical observations as PMS stars possess little or no circumstellar emission at optical wavelengths. Owing to large off axis beam of Chandra, we
checked these likely members for the counterparts of our optical dataset with a less
stringent matching radius of 3 arcsec. In a few cases, where there were more than one source within
the matching radius, we considered the closest one as the best match \citep[e.g.,][]{pan17}. This search revealed
possible optical counterparts to 257 (72 \%) sources in the region A.
 
\section{Results \& Discussion}
\begin{figure}
\centering
\includegraphics[scale = 0.4, trim = 0 0 0 10, clip]{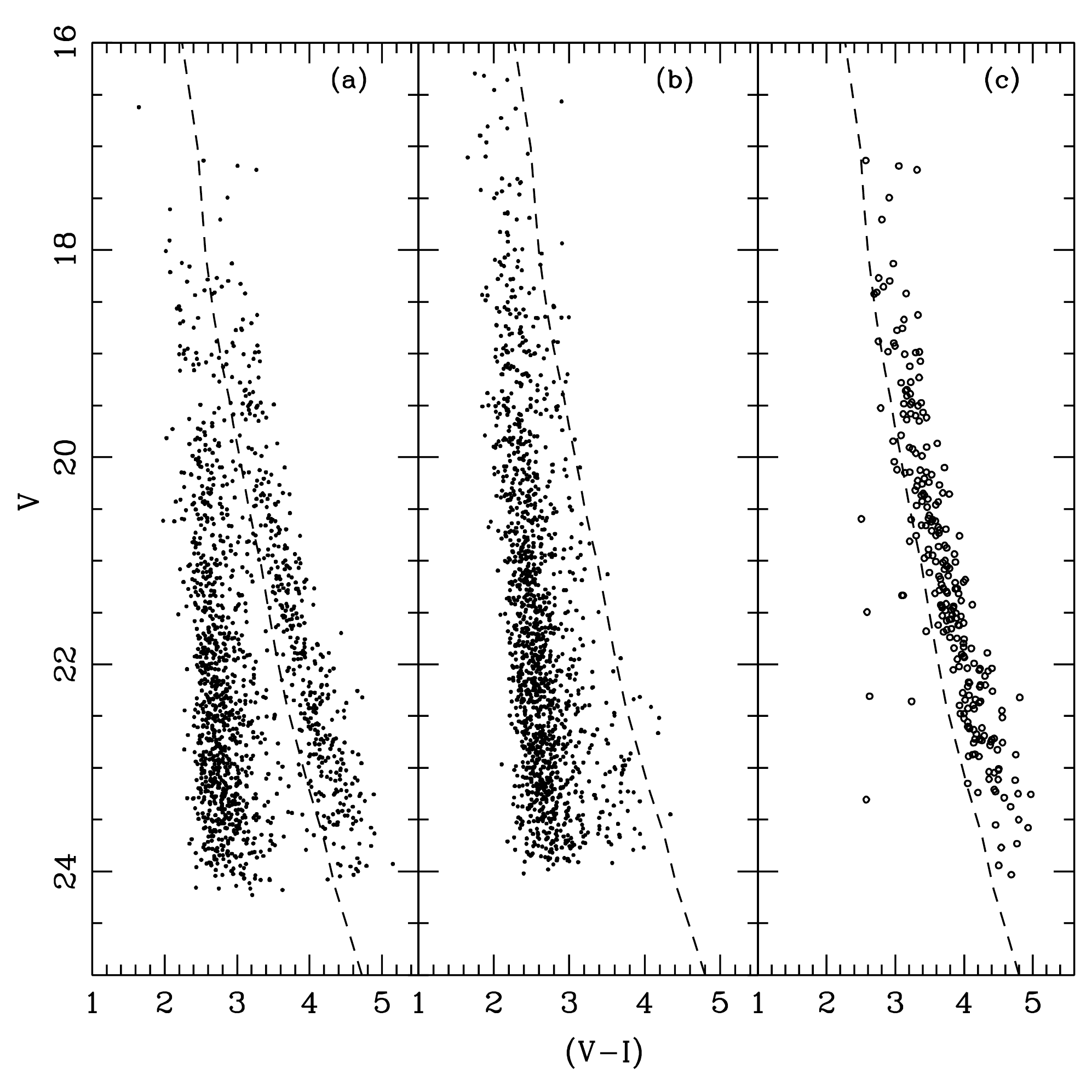}
\caption{Color-magnitude diagrams (CMDs) for the stars in (a) region  A, and (b) the control field. 
 The dashed line is drawn to tentatively separate the PMS population from the field stars. 
(c) CMD for the known candidate YSOs from \citet{get17}. }
\label{cmd}
\end{figure}

\subsection{Identification of PMS members of the cluster}

The color magnitude diagram (CMD) is an ideal tool to study the evolutionary stages of the member stars of the cluster. 
However, CMDs can be contaminated severely by the field population along the line of sight \citep[e.g.,][]{sung10,da10,cha11,sung13,pan13a}. The 
cluster population can be distinguished from the general Galactic population along the line of sight through a comparison of the CMDs of the 
cluster region and the control field \citep[e.g.,][]{bran08,shar08,gen11,hay15}. In Fig \ref{cmd}a and \ref{cmd}b, we show the  $V$ vs. ($V$ - $I$) CMDs for the stars in the 
cluster region and control field,
respectively. As can be seen, the
cluster PMS population clearly stands out in the cluster CMD compared
to the control field CMD. There is a clear color gap between the likely 
PMS sources of the cluster and that of the field stars 
towards that direction. 
Comparing the cluster and field regions, we consider those stars that lie 
right of the dashed curve (see Fig. \ref{cmd}a) as the PMS population of the cluster. 
To further confirm the PMS zone, we compare the distribution of the previously
identified YSOs (i.e, YSOs identified by using {\it Spitzer} and X-ray observations; see Sec. 2.3) on the 
CMD (see Fig. \ref{cmd}c) to the distribution of the field stars. In doing so, 
we found that most of the YSOs are actually located right of the dashed line,
reinforcing our hypothesis that the contamination due to field stars in the PMS population 
of the Be~59 region is insignificant. Comparing the number of field stars to that of stars located right of the dashed line, the level of contamination by field stars in our PMS sample is less than 3\%. 

\subsection{Extinction measurement}
\begin{figure*}
\centering
\includegraphics[width=7cm]{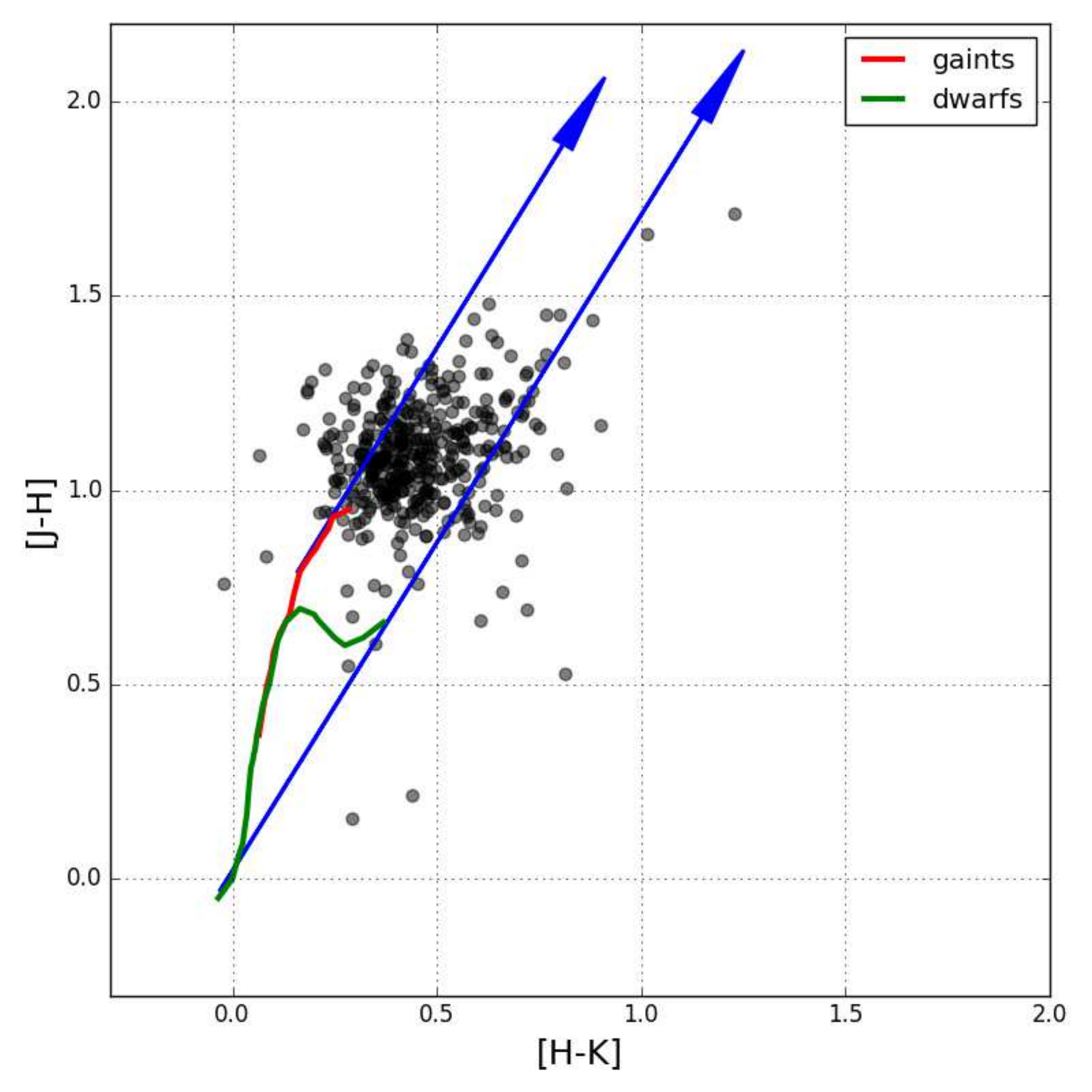}
\includegraphics[width=7cm]{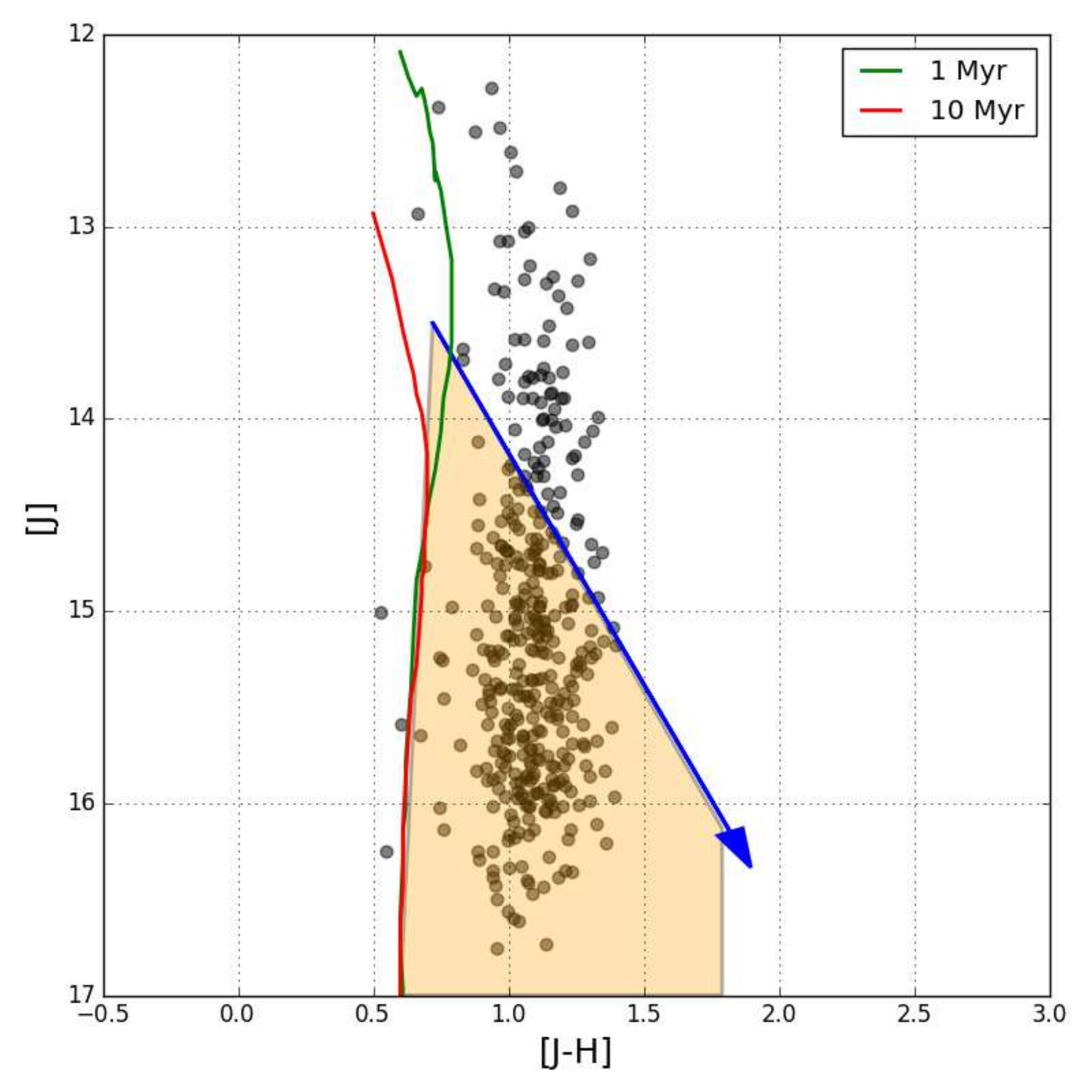}
\includegraphics[width=7cm]{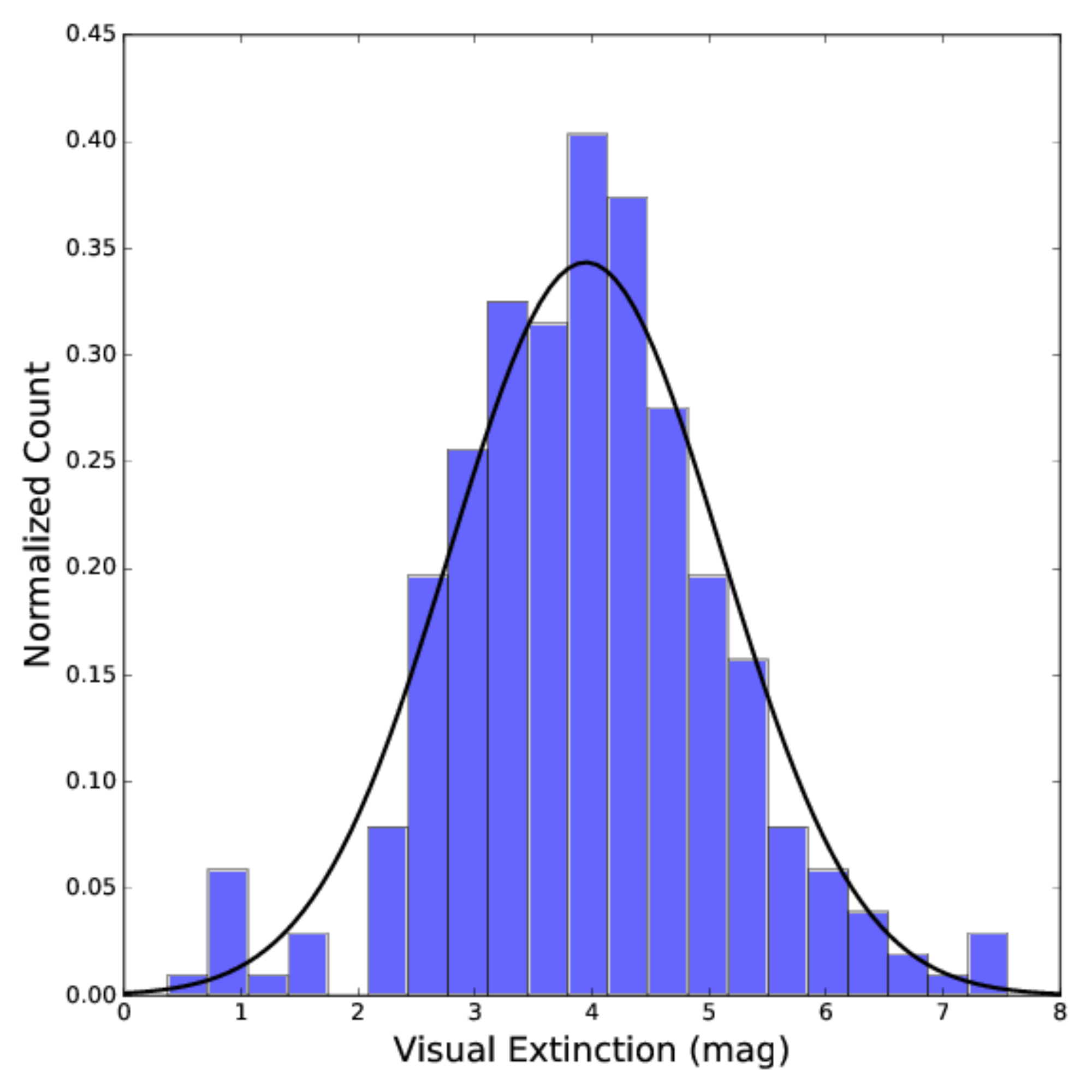}
\caption{Upper left panel : 2MASS (J - H) vs. (H - K) color-color diagram of the PMS stars. 
The red solid line show typical colors of giants from \citet{bes88}. 
The solid green curve shows the intrinsic colors of the MS stars from \citet{bes88}. The blue arrows indicate the reddening  
vectors according to the extinction laws of \citet{coh81}. Upper right panel:  J vs. (J - H) color-magnitude diagram of the PMS stars without IR-excess. 
Over-plotted are 1 and 10 Myr isochrones from \citet{bar98} for the adopted distance (1 kpc) and zero reddening. The shaded area represents the sample selected for extinction
measurements. The arrow is the reddening vector according to the extinction law of
 \citet{coh81}. Bottom panel: Distribution of the derived visual extinction of the PMS population. 
The solid curve represents the Gaussian fit to the distribution.}
\label{ext}
\end{figure*}

Previous studies have determined reddening ($E$($B$ - $V$)) of Be~59 in the range of 1.4 to 1.8 mag from the optical measurements of a few bright stars \citep{pan08,maj08}. 
Since, the current data-set extends substantially deeper than the previous surveys and the PMS content is clearly identified. Here, we utilize the near-infrared colors of the low-mass content of the cluster
to derive its extinction, since the near-infrared colors of low-mass stars are less sensitive to the change in spectral types. 
To do so, we selected all the PMS sources that lie to the 
right of the dashed line in the optical CMD (see Fig. \ref{cmd}a), then searched for their NIR counterparts within a 1$^{\prime\prime}$ and followed the same procedure as outlined in Sec. 2.3. The NIR colors of 
these sources are shown in Fig. \ref{ext} (upper left panel). 
As can be seen from this (J - H) vs. (H - K) color-color diagram, a majority of them are located in the reddened MS zone 
(i.e, the zone between the two reddening vectors), and only a few sources show characteristics of infrared excess 
(i.e, those sources located beyond the right reddening vector), indicating that the likely excess 
emission due to the circumstellar disk is less or minimum for a majority of the sources \citep[for details on NIR color-color diagram, see][and references theirin]{ojha09}. 
The fluxes in the redder infrared bands may be contaminated by the infrared excess emission arising from the circumstellar disk. However, it is 
generally considered that among near-infrared bands the effects of the excess emission in J-band is minimum 
\citep[e.g.,][]{mey97a,luh03b}. 
Even though only a few PMS sources 
show infrared excess in the (J - H) vs. (H - K) color-color diagram, we select only those sources that do not show characteristics of infrared excess in order 
to minimize the effect of excess emission on extinction. 
Figure \ref{ext} (middle panel), shows the J vs. (J - H) diagram of these  PMS sources. Over-plotted solid curves are the 1 Myr and 10 Myr isochrones from
\citet{bar98} with zero reddening and at the adopted distance (1 kpc). In both the plots the isochrones and photometric magnitudes are converted into the CIT system by using 
the transformation equations from \citet{car01}. As can be seen from the figure, at the low-mass end the 
PMS isochrones are almost vertical in the J vs. (J - H) diagram, particularly below J=14 mag. 
As the color change for the  low-mass PMS objects (J $>$ 14 mag) due to age variations within the expected age range of 1 - 10 Myr is minor, so 
de-reddening is possible with a unique solution for each star. To get better statistics on the extinction, we use a slightly relaxed criterion by selecting those sources 
that lie below the reddening vector (shown as an arrow) originating at J = 13.5 mag (instead of J = 14 mag) and (J - H) = 0.68 mag, 
i.e, the sources within the shaded
area in the plot. We then derived the color excess  of these selected sources
by tracing them back to the vertical line of the plot along the reddening vector,
then converted the color excess to the A$_V$ (=9.1 $\times$ E(J - H), for R$_V$=3.1)
value using the reddening law of \citet{coh81} which is in the CIT system.

The distribution of the derived visual extinction values is shown in Fig. \ref{ext} (lower panel). 
The peak and dispersion of the visual extinction distribution as estimated from a Gaussian fit are 4.0 mag and 1.2 mag, 
respectively. This mean visual extinction corresponds to an $E$($B$ - $V$)=1.3 mag. 
This is in very good agreement with the extinction estimation based on the bright stars by previous authors.
For the further analyses, we use A$_V$= 4 mag as the mean visual extinction of the region. 

\subsection{Age/Mass of the YSOs in the cluster region}
\begin{figure*}
\centering
\includegraphics[scale = 0.44, trim = 0 0 0 0, clip]{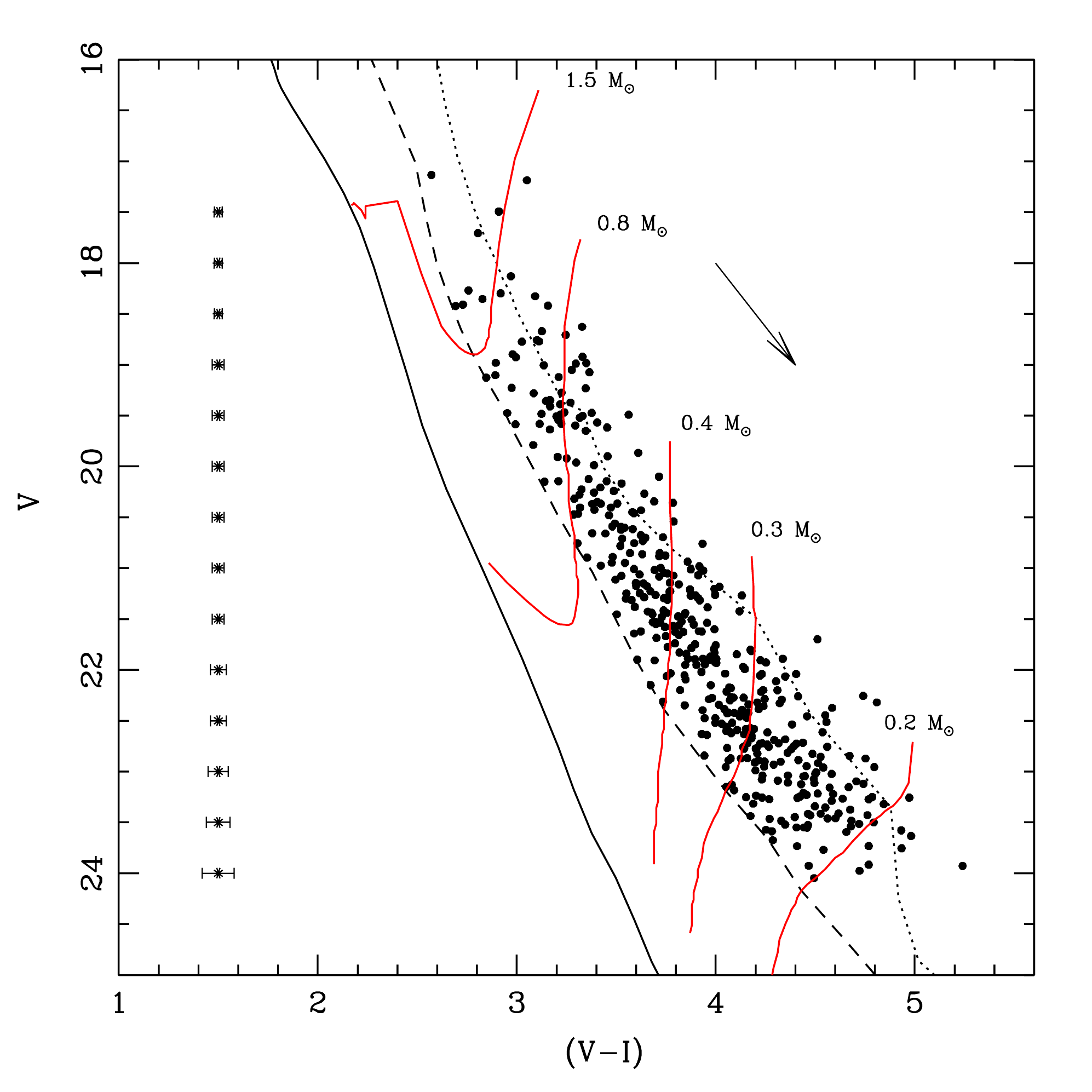}
\caption{Color-magnitude diagram for the PMS sources in the present work. 
The 2 Myr MS (continuous curve) by \citet{gir02} as well as the PMS 
isochrones for 1 and 5 Myr (dotted and dashed curves, respectively) having solar metallicity by \citet{sie00} 
are also plotted. The isochrones are shifted for the adopted distance (1 kpc) and reddening ($E$($B$ - $V$)=1.3 mag). The arrow represents the reddening vector of A$_V$=1 mag. The average error in ($V$ - $I$)
color as a function of magnitude is shown on the left-hand side. \\
 }
\label{ysocmd}
\end{figure*}

 The ages of the massive stars are difficult to estimate as they quickly arrive on the main sequence while 
the low-mass stars are still on their PMS tracks. Thus  ages of young clusters are typically derived from post-main-sequence evolutionary tracks 
for the massive members showing significant evolution, or based on theoretical PMS isochrones for low-mass contracting population. Here, we rely on a sample 
of young stars (the PMS sources
identified in Sec. 3.1) to derive the age of Be~59. 

Fig. \ref{ysocmd} represents the optical $V$/($V$ -- $I$) CMD for our PMS stars in the region A. 
The 2 Myr main sequence (MS) by \citet{gir02} as well as the PMS 
isochrones for 1 and 5 Myr having the solar metallicity by \citet{sie00} 
are also plotted\footnote{Here, we used \citet{sie00} isochrones because the evolutionary models of \citet{bar98} is limited to 1 Myr and lack isochrones for masses above 1.3 M$_\odot$.
Consequently, the median age of the cluster can only be constrained to $<$1 Myr. On the other hand, \citet{sie00} 
models extend to younger ages than those of \citet{bar98}, thus provide
us with additional information on the age of the cluster population,
as well as means to compare our results with previous results of Be~59, where the authors have used the Siess models to 
derive the membership, age, and mass function \citep[e.g.,][]{pan08,esw12}.
}. 
The isochrones are shifted for the adopted distance (1 kpc) and reddening, $E$($B$ - $V$) = 1.3 mag. 
To determine the age and mass of each star, we downloaded a large number of 
PMS isochrones from \citet{sie00} for 0.1 to 10 Myr with an interval of 0.1 Myr 
from their web interface. We then interpolated each isochrone to make them to be more continuous curves.  Finally, we assign the age and mass to a star corresponding to the closest isochrone on the CMD. The age and mass of the YSOs are given in Table \ref{tab4}. Only a portion of Table \ref{tab4} is given here and the complete table is available in
the electronic version.

\begin{figure}
\centering{
\includegraphics[width=6cm]{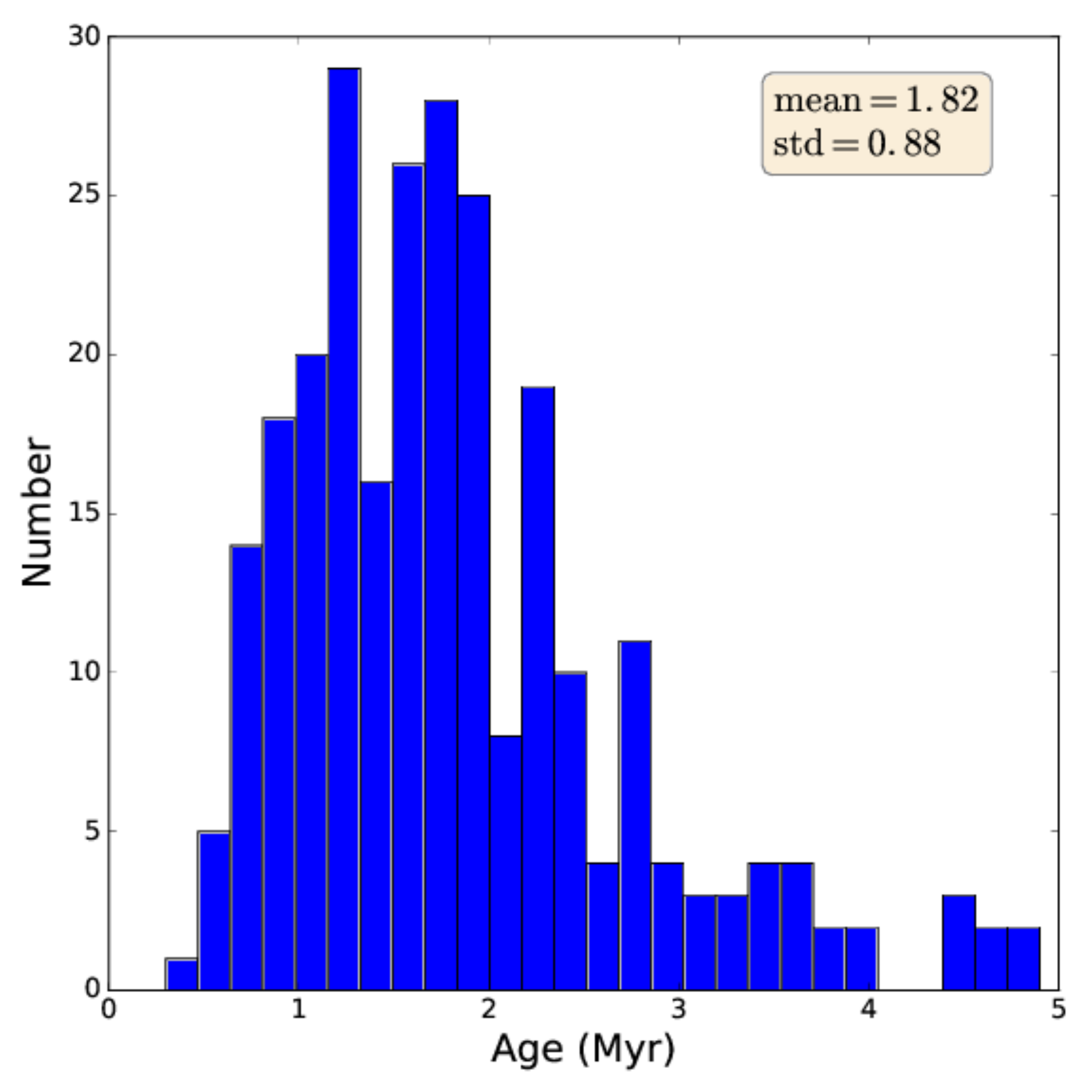}}
\caption{Age distribution for the PMS stars with mass $>$0.3 M$_\odot$.\\}
\label{ageyso}
\end{figure}

The age distribution derived for the PMS sources is shown in Figure \ref{ageyso}. As one can see from the figure, the 
distribution shows a scatter in age primarily between 0.5 to 5 Myr. The mean age of the cluster population is $\sim$ 1.8 $\pm$ 0.9 Myr, which is in good agreement with that obtained by \citet{pan08}.

In the case of young clusters this type of age spread 
is common and may be due to the combined effect of differential reddening,
circumstellar disk extinction, binarity, variability and/or different
evolutionary stages of the cluster members. 
Since, in the $V$ vs. ($V$ - $I$) CMD, the reddening vector is nearly parallel to the isochrones, a small differential reddening 
(i.e., $\Delta$A$_V$ = 1 mag in the present case) 
may not significantly affect the age estimation. 
Though the circumstellar disk may affect the age and mass
determination of YSOs in IR CMDs, its effect should be
minimal in optically visible sources such as sources with 
optically thin disks \citep[e.g.,][]{jos13}. 
In fact as one can see from Fig. \ref{ext} (upper panel), most of the PMS sources do not possess optically thick disk at NIR bands. 
A binary system appears brighter and, consequently will be assigned a younger age 
on the basis of the CMD. For example, in the case of equal-mass binaries, a cluster is expected to show a sequence which is shifted by 0.75 mag upwards. The effect of binarity can be 
more prominent in somewhat evolved clusters ($\sim$ 10 Myr)
in which many member stars are in Henyey track and so the isochrones are closer 
to each other, whereas in the case of very young
clusters the isochrones are more separated, hence the effect of binarity
on age estimation will be less compared to the natural
age spread of the stars in the clusters \citep[e.g.,][]{sic05,pan17}. 
Similarly though, most of the young stars (T-Tauri stars) show variability at optical bands \citep[e.g.,][]{her94,rod10,lat15}, however, 
as demonstrated by \cite{bur05}, due to variability the location of 
objects in the CMD tend to move parallel to the isochrones, resulting in little effect
on the age estimation \citep[see also discussion in][]{sch12}.

Despite of the low differential reddening and presence
of large number of disk-less sources in the cluster,
some degree of age spread due to the combination of
the above factors is expected, but since the observed
spread in color among the PMS members is larger than
their typical uncertainties due to photometric colors
(see Fig. \ref{ysocmd}), we considered that PMS stars are likely at different
evolutionary stages. 

We consider that the errors associated with the determination of age and mass are mainly of two kinds; random
errors in photometry, and systematic errors due to different
theoretical evolutionary tracks. We estimated the effect of
random errors by propagating them to the observed estimations
of $V$, ($V$-$I$) and E($V$-$I$) by assuming a normal
error distribution and by using Monte Carlo simulations \citep[see
e.g.,][]{cha09}. Since we have used the same evolutionary model \citep[i.e.,][]{sie00}
to calculate the age/mass for all the PMS stars, the results should not be affected by systematic errors. 
\begin{table*}
\centering
\caption{Magnitudes, mass and age of the PMS stars. Complete table is available in an electronic form. }
\begin{tabular}{ccccccc}
\hline
\hline
Id     &RA (J2000) & DEC (J2000)&             V $\pm$ eV & I $\pm$ eI             & Mass $\pm$ error in Mass (M$_\odot$)& Age $\pm$ error Age (Myr) \\
\hline
\hline
   1    &0.668875 & 67.343697 &20.90 $\pm$  0.02  &  17.54  $\pm$   0.02   &0.73 $\pm$ 0.16  &   4.50 $\pm$ 0.33  \\
   2    &0.411083 & 67.346443 &19.57 $\pm$  0.02  &  16.16  $\pm$   0.02   &0.62 $\pm$ 0.10  &  0.76 $\pm$ 0.05 \\
   3    &0.461458 & 67.347725 & 22.34  $\pm$  0.02  &  18.33  $\pm$   0.03   &0.34 $\pm$ 0.04  &  2.40 $\pm$ 0.10       \\
... &...&...&...&...&...&...\\
\hline
\end{tabular}
\label{tab4}
\end{table*}

\subsection {Initial Mass Function and Mass  of the Cluster}
The IMF is the distribution of the masses of stars 
at the time of a star formation event. Young clusters are preferred sites for IMF studies as 
their mass functions (MFs) can be considered as IMFs, since they are too young to loose 
a significant number of members either due to dynamical evolution or stellar evolution.
The variation of the IMF gives clues to the
physical conditions of star formation processes \citep[e.g.,][]{bate09}. The IMF
is defined as the number of stars per unit logarithmic mass
interval, and is generally represented by a power law with the 
slope, \\ 
$\Gamma$ = {{ $\rm {d}$  log $N$(log $m$)}/{$\rm{d}$ log $m$}},\\
 where $N$(log $m$) 
is the  number of stars per  unit logarithmic mass interval. 
The classical  value derived  by \citet{sal55} 
is $\Gamma = -1.35$ for the mass range  $0.4  < M/M_{\odot} \le 10$.

\begin{figure}
\centering
\includegraphics[scale = .4, trim = 5 5 5  5, clip]{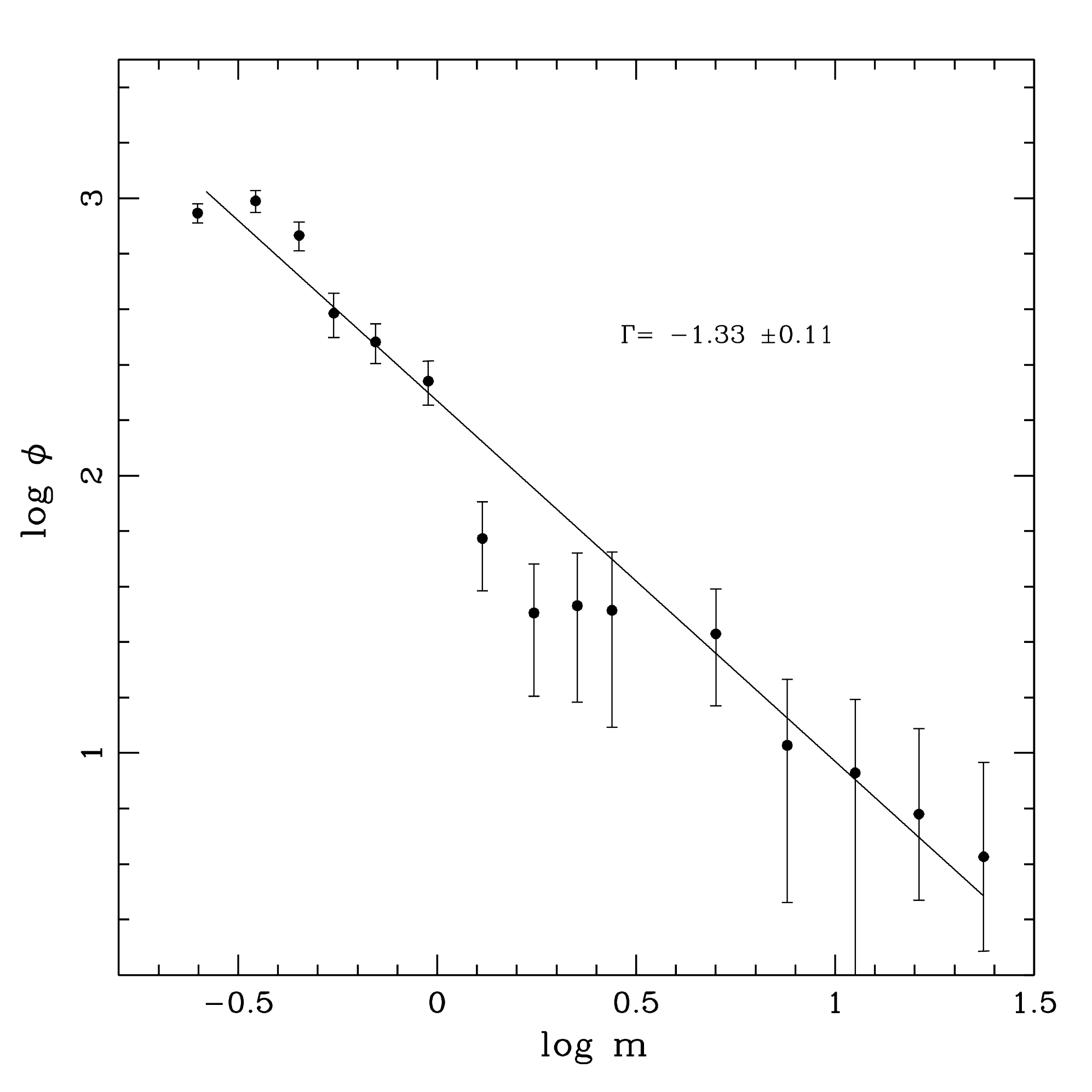}
\caption{ The IMF in the region derived from the optical data. 
The $\phi$ represents $N$/{\rm{d} log $m$ }. The error bars 
represent the $\pm$$\sqrt{N}$ errors. The continuous line shows the weighted least-squares 
fit in the mass ranges described in the text. The value of the
slope obtained is given in the figure.
}
\label{mf1}
\end{figure}

Here, we used the optical CMD to count the number of stars in different mass bins, which is shown in Fig. \ref{ysocmd} along with isochrones and evolutionary tracks. 
To make corrections to the data sample for incompleteness which may be due to various reasons, e.g., crowding of the stars, we used the ADDSTAR routine of $DAOPHOT$-$II$ to determine 
the CF as described in Sec. 2.1. 
 The IMF of the PMS stars was obtained by counting the 
number of stars in different mass bins and correcting for incompleteness \citep[for details see][]{pan08,jos08}. 
Fig. \ref{mf1} shows the MF of the cluster region A in the mass range 
0.2 $\le$ M/$M_\odot$ $\le$ 28. We note our data is sensitive to the mass-range 0.2 - 1.5 M$_\sun$.
Since most of the bright stars ($V$ $<$ 18 mag with the corresponding mass $>$ 1.5 M$_\sun$) were 
saturated in our observations, the data for higher masses ($>$ 1.5 M$_\odot$) have been taken from \citet{pan08}. 

As can be seen from the figure, the mass function rises up to 0.3 $M_\odot$. Deep infrared observations will be more efficient to 
shed light on the sub-stellar population of the cluster and its IMF. Nonetheless, 
we note, the low-mass IMF has been investigated for a number of
other star-forming regions, and in some cases the results are similar to the IMF derived here. For example, a
variety of photometric and spectroscopic surveys have been completed in the extremely dense Trapezium cluster and the 
intermediate density cluster IC~348 by multiple authors \citep{hill00,mue02,mue03,sle04,luc05}. All of them 
found an increasing mass functions to a maximum at $0.1-0.3$ $M_\odot$ before declining 
in the brown dwarf regime. These results are very different from the mass function 
derived for the Taurus star-forming region (characterized by its low gas and stellar densities), 
which peaks at 0.8 M$_\odot$ \citep{bri02,luh03a,luh04}.

A single slope for the IMF in the mass range 
0.2 $\le$ M/$M_\odot$ $\le$ 28 can be fitted with $\Gamma$ = -1.33 $\pm$ 0.11, which 
is nearly equal to the  Salpeter value (-1.35). If we fit a single slope to the data in the 
mass range 0.2 $\le$ M/$M_\odot$ $\le$ 1.5, the resultant $\Gamma$ value is -1.23 $\pm$ 0.26, which 
is  comparable to the Salpeter value. We note, \citet{pan08} estimated a 
slope of -1.01 in the mass range 2.5 - 28 $M_\odot$, which could be due 
to combination of 
low statistics of stars at high-mass end, small mass range  and different area 
used by them. 

It is difficult to make an accurate estimate of the cluster mass without knowing all the low-mass members.
We can however make a relatively good estimate from the present data as they reach down to 0.2 M$_\odot$. 
Most of the observations of the high-mass SFRs in our Galaxy
are consistent with an IMF that declines below 0.2 M$_\odot$ \citep[e.g.,][]{luh07,and08,oli09,sung10,nei15,jos17}. 
Thus the contribution below it is not expected to substantially alter the estimate of the total mass of the cluster if we assume that the cluster contains stars down to 0.08 M$_\odot$. 
Adopting -1.33 as the slope of the mass function, we estimate the total mass of the cluster as $\sim$ 950 M$_\odot$ by integrating the mass function
between 0.2 and 28 M$_\odot$. The 
total number of stars in the cluster down to 0.2 M$_\odot$ is estimated to be $\sim$ 1500.
The stellar density profile of the cluster based on the optical data suggests that most of the stars are located within 5 - 8 arcmin radial distance from the center of the cluster \citep{pan08}. Although our data represent majority of the cluster members, however, we emphasize that the estimated mass is a lower-limit to the actual total mass of the cluster and 
only represents the mass within the central part of radius $\sim$ 4 arcmin ($\sim$ 1.2 pc). 

\subsection{Mass Segregation}
\begin{figure*}
\centering{
\includegraphics[width=8cm]{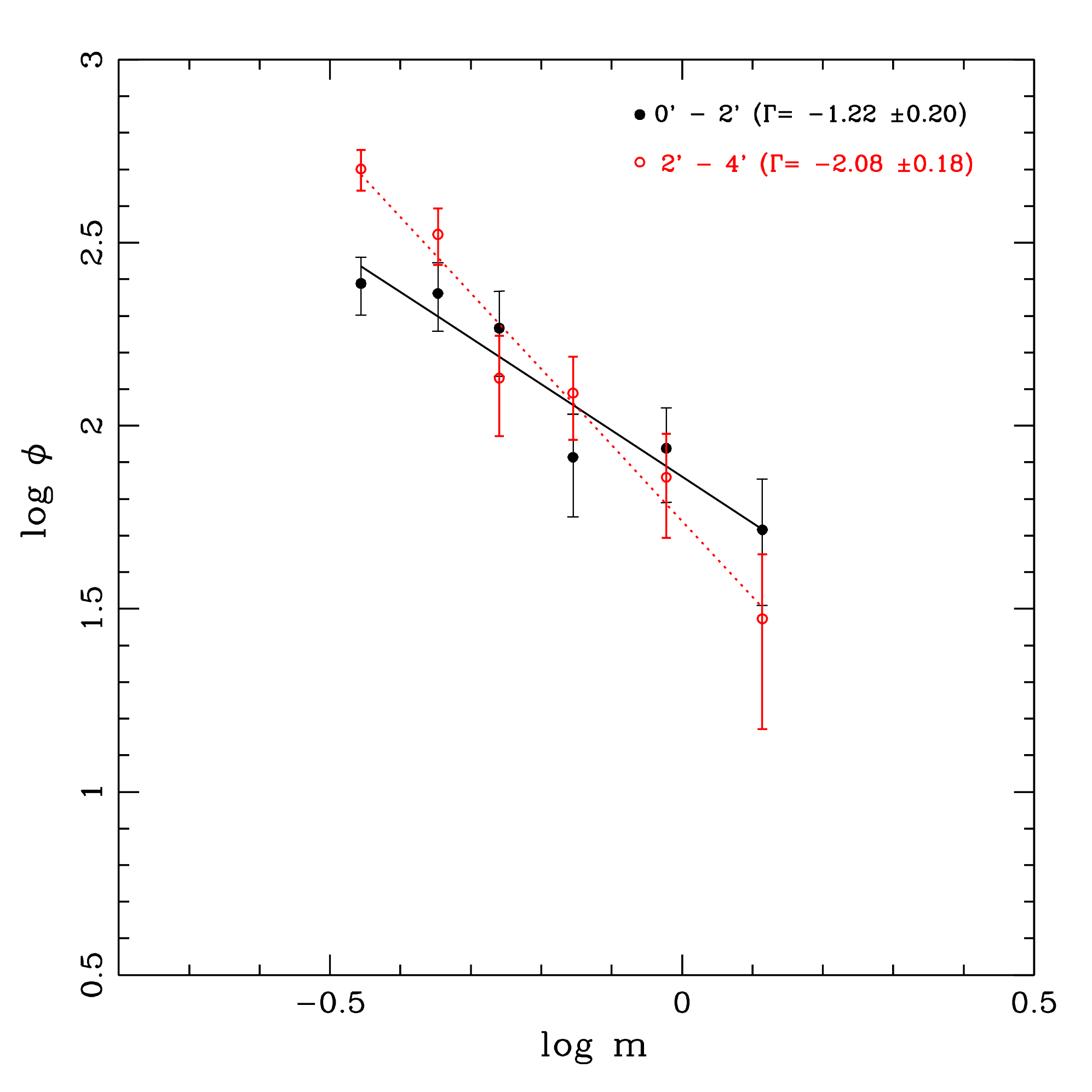}
}
\caption{IMFs for the low-mass stars lying within 0-2 arcmin and 2-4 arcmin from the cluster center. The $\phi$ represents $N$/{\rm{d} log $m$ }. The solid lines show the weighted least square fit to the data.}
\label{ms}
\end{figure*}

The concentration of massive cluster members at the center
and lower mass members at larger radii is
commonly referred as mass-segregation in clusters.
Whether mass segregation is primordial or dynamical, is an important constraint on theories of massive star formation as well as 
cluster formation and evolution. The competitive accretion theory \citep{bonn01,bonn06} suggests that protostars 
in the dense central regions of a young star cluster can accrete more material than those in the outskirts, therefore 
primordial mass segregation would be a natural outcome of massive star formation. However, mass segregation can 
occur dynamically \citep{kro01,mcm07}. In this scenario massive stars form elsewhere in the cluster but 
sink to the cluster center through dynamical interaction with low-mass stars. The dynamical evolution drives the
system towards equipartition, the natural result of this being that the lower-mass stars attain higher velocities, hence 
occupy larger orbits around the cluster centre. In turn, the higher-mass stars will sink towards the cluster centre.

In Be~59, as discussed in \citet{pan08}, the high-mass stars are more 
centrally concentrated than their lower mass siblings.
Although, the issue of mass segregation has been extensively investigated in clusters by focusing on 
high-mass stars \citep[e.g.,][]{chen07,pang13,lim13,wang14}, the possibility of differential mass segregation among low-mass stars has been
rarely discussed \citep[e.g.,][]{and11}. Some  theories concerning the formation of very low-mass stars suggest such objects 
may have a greater velocity dispersion than their higher mass siblings and hence be less spatially concentrated 
in a cluster \citep[e.g.,][]{ste03}. 

In this work, we investigated the mass segregation by looking at the radial variation of the IMF of low-mass stars. To do so,
we divided the cluster into two radial bins of width 2 arcmin and derived the IMFs using the stars in the mass range 
0.3 - 1.5  M$_\odot$. Figure \ref{ms} shows the IMFs of the cluster in these two bins. As can be seen from the figure,
the slope of the IMF in the inner bin (i.e., between 0-2 arcmin) is clearly flatter than the outer bin (i.e., between 2-4 arcmin),
although the fluctuation of the data points is rather large due to the low statistics of stars. 
Since the mass function slopes  differ from each other by more than 4$\sigma$ uncertainty, where $\sigma$ is the error associated to the slope, we view this trend as an
evidence for mass segregation in the cluster. 

The time-scale for the onset of significant dynamical mass segregation is 
comparable to the dynamical relaxation time of the cluster.  
We estimate the dynamical relaxation time $t_{relax}$ for the cluster: \\
$t_{relax} = \frac{N}{8
\ln N}\times t_{cross}$,\\
where $t_{cross} = \frac{R} {\sigma_v}$ is the crossing time, 
R is the size of the cluster, N is the total number of stars and 
$\sigma_v$ is the velocity dispersion of the stars \citep{bonn98}.
From the CO observation of Be~59 \citep{liu88}, the velocity dispersion of the molecular gas is $\sim$ 1.4 km s$^{-1}$. 
In the absence of kinematic information of the stellar members of the 
cluster we use the velocity dispersion of CO gas to roughly estimate 
the relaxation time. Considering $R\sim$ 2.6 
pc, $\sigma_v$ $\sim$ 1.4 km s$^{-1}$ and N $\sim$ 1500,
we obtain $t_{relax}\sim 4.5$ Myr for Be~59. As the cluster age is $\sim$ 1.8 Myr, 
no significant mass segregation is expected from two-body dynamical interactions.  
The cluster still 
retains a significant amount of molecular gas and cold dust components \citep[see discussion in][]{pan08,yan92} that are usually thought to 
impede the process of dynamical mass segregation. From the above evidences, we  
argue that the mass segregation is likely primordial, although precise kinematic measurements of stellar members would shed more light on this issue.
 
\subsection{Comparison to other nearby clusters}

\begin{figure}
\centering
\includegraphics[width=8cm]{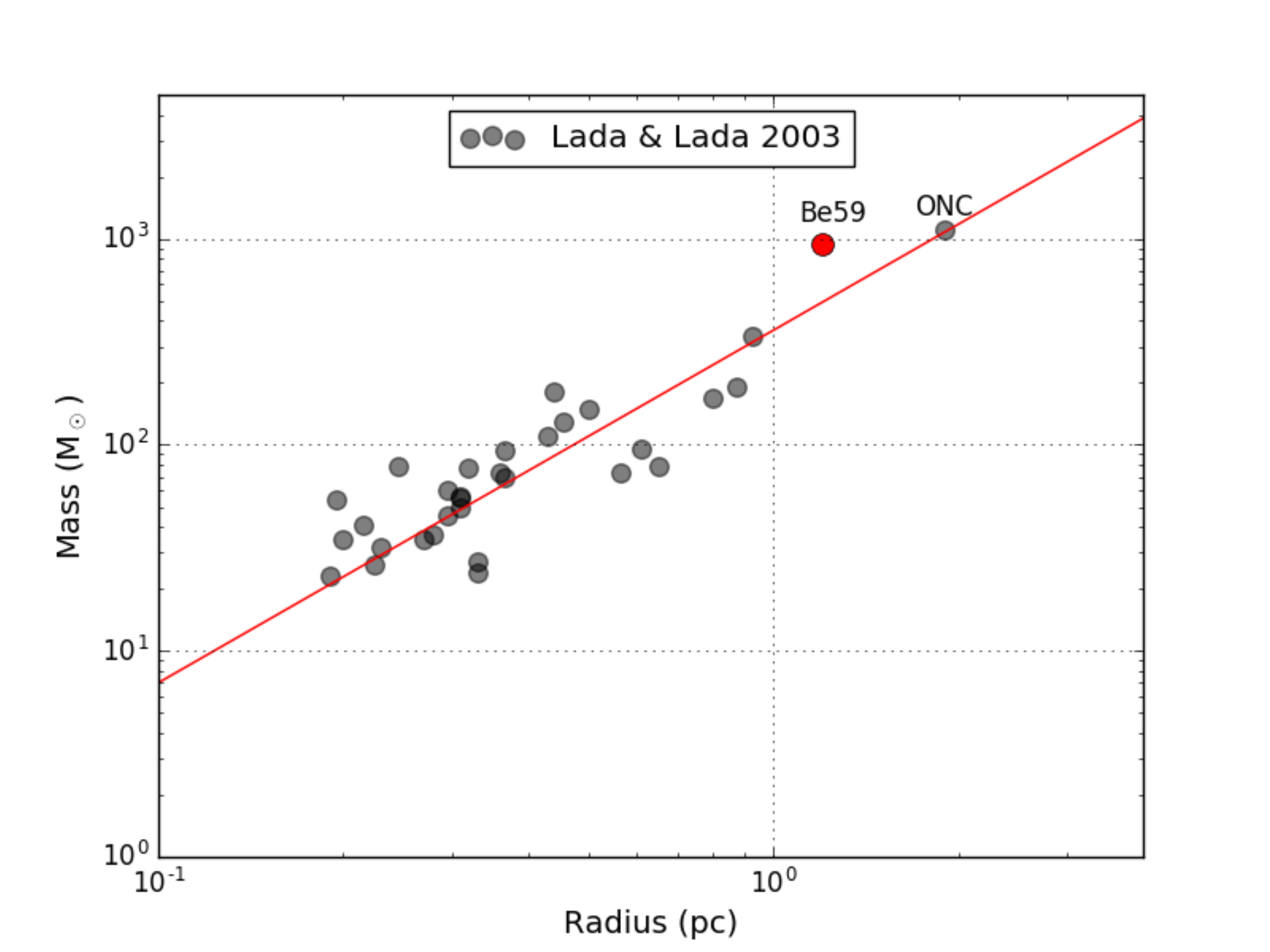}
\caption{Comparison of Be~59 with the other nearby clusters. The cluster sample is from \citet{lada03}. 
The location of Be~59 and ONC is marked. }
\label{cmr}
\end{figure}

With the total mass of $\sim$ 1000 M$_\odot$, the age $\sim$ 2 Myr and the stellar population containing early-mid O-type stars, Be~59 presents
a nearby, rich and partially embedded young cluster. Therefore, it is worthwhile to compare the properties of Be~59 
with those of the nearby embedded clusters in our Galaxy.
Figure \ref{cmr} shows the data sample (black dots) of embedded clusters from the \citet{lada03} with known sizes and masses. 
This data set is limited to embedded clusters within 2 kpc from the Sun, and, like Be~59, they are far 
enough away from the Galactic centre and the spiral arms.
In Fig. \ref{cmr} the solid line shows the relation, 
mass $\propto$ radius$^{1.7}$, obtained by \citet{pfl11} by using the \citet{lada03} sample. As can be seen, Be~59 follows the above mass-radius relation well. 
Interestingly it resembles Orion Nebula Cluster (ONC). 
ONC is one of the nearest young clusters with approximately 1000 M$_\odot$ of stars \citep{hil97,da10}. 
The most massive star of Trapezium, $\theta$ Ori C, is of spectral type O7V \citep{stah08}
The second massive is of O9.5V, and the rest two are of B0-B1. 
 \citet{regg11} calculated the age of ONC based on the \citet{sie00} evolutionary
models and found a mean age of around 2.2 Myr \citep[see also][who reached a nearly smiliar conclusion]{jeff11}. 

Compared to ONC, Be~59 contains $\sim$ 1500 stars, with the total stellar mass of $\sim$ 1000 M$_\odot$ and a mean age of $\sim$ 
1.8 Myr. Also  the spectral types of the four bright members (lying within 0.24 pc from the center) in the Be~59 are 
similar to those of the Trapezium system (see Table \ref{trap}). So we argue that Be~59 is an ONC-like cluster located in the 
the Galaxy at 
the distance of $\sim$ 1 kpc.

\begin{table*}
\caption{Four massive members of the cluster Be~59} 
\begin{tabular}{|p{1.0in}|p{0.4in}|p{0.4in}|p{0.4in}|p{1.7in}|}
\hline
Name    & distance from the cluster center (in pc)& $V$ mag &Spectral Type&Reference\\
\hline
BD +66 1675     & 0.17      &  9.08 & O7&\citet{maj08} \\
BD +66 1674     & 0.24      &  9.60  & B0&\citet{maj08}\\
LS I +67 9      & 0.02     &  11.30 & B0.5&\citet{maj08}\\
TYC 4026-0424-1	& 0.23      & 11.81 & O9&\citet{mai16}\\
\hline
\end{tabular}
\label{trap}
\end{table*}

\subsection{Clues on cluster formation}

\begin{figure*}
\centering{
\includegraphics[width=8cm]{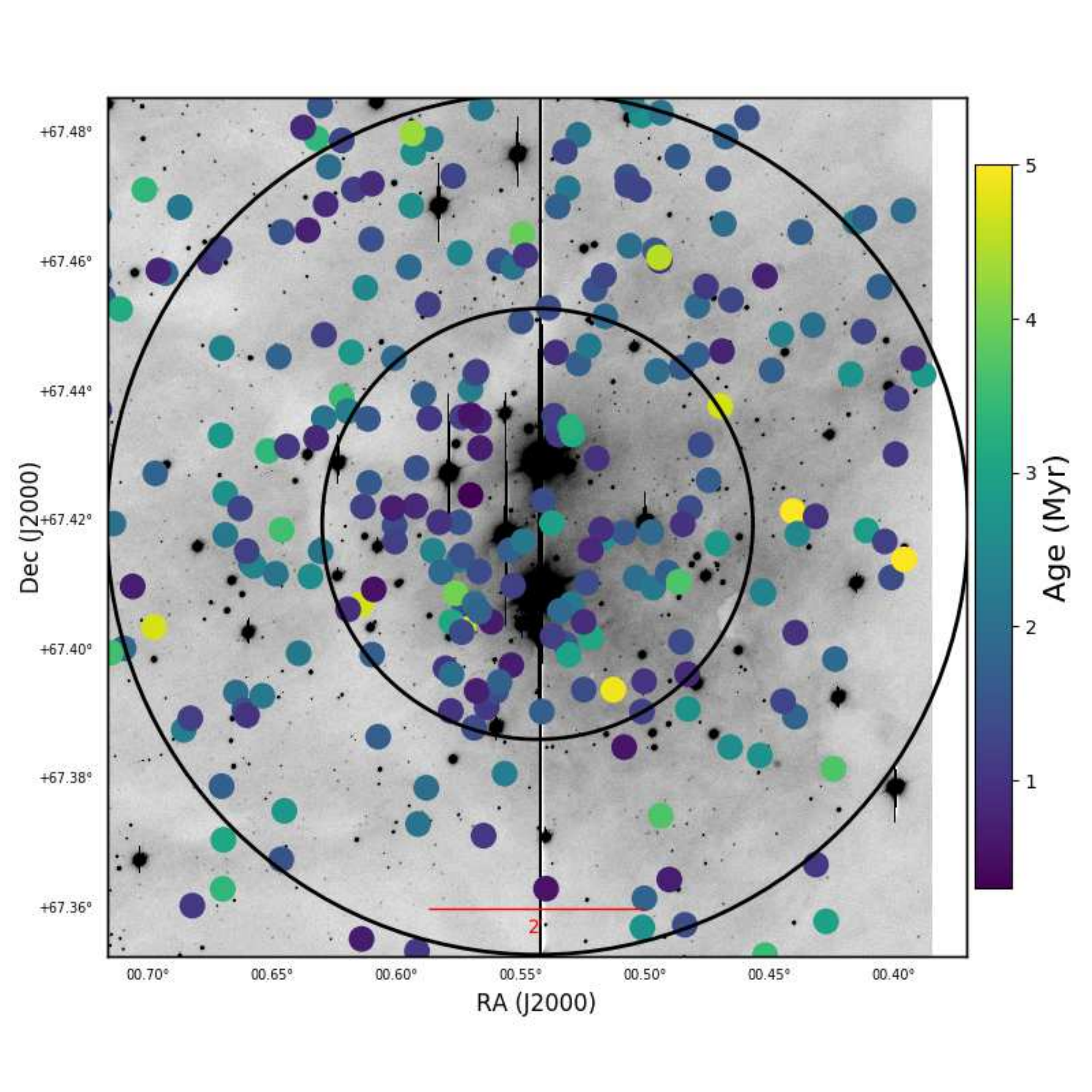}
\includegraphics[width=8cm]{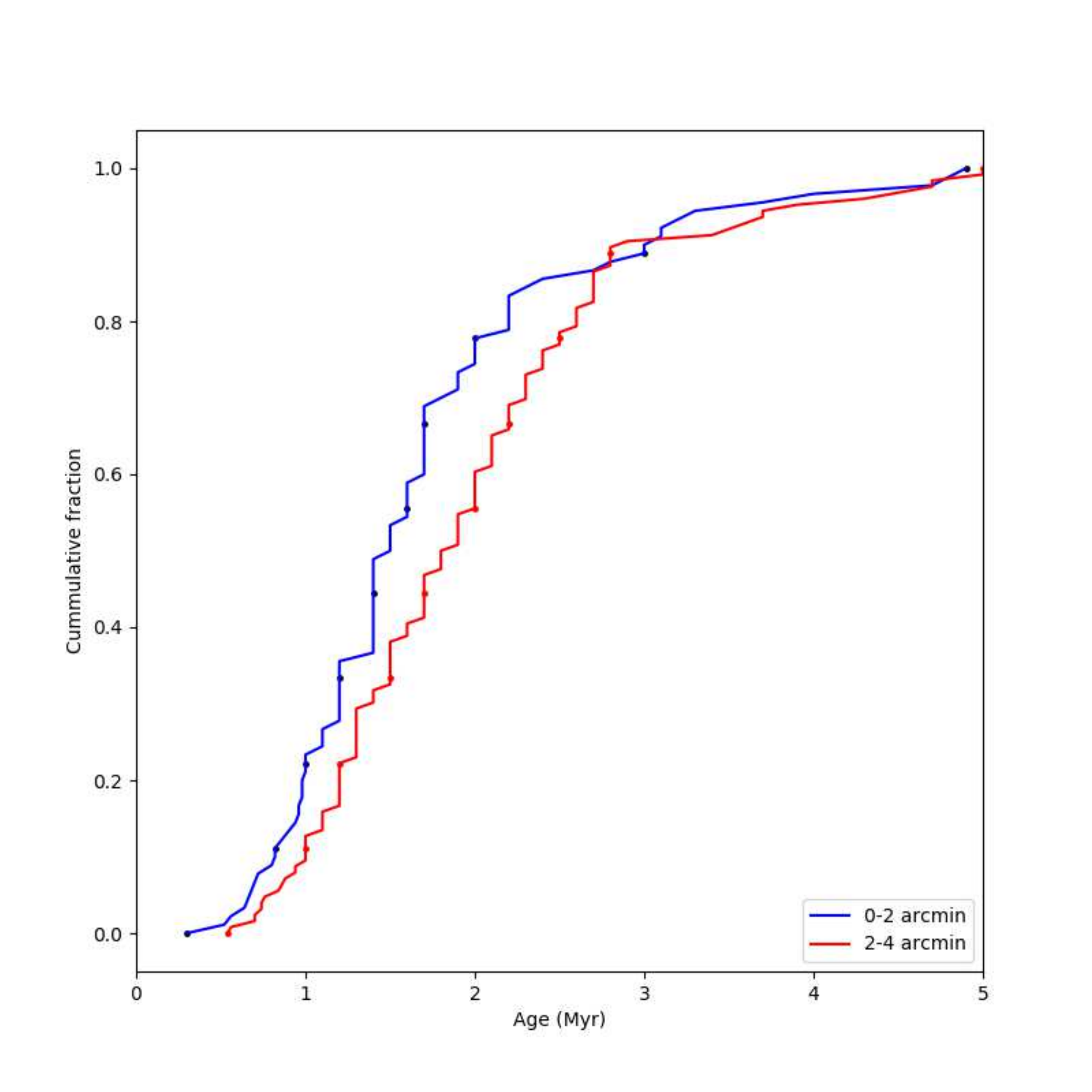}}
\caption{Left panel: Age distribution of low-mass stars. The inner and outer circles present radii of 2 arcmin and 4 arcmin, respectively. 
Right panel: KS test on the 
age distribution of the stars within 0 - 2 arcmin and 2 - 4 arcmin from the cluster center. }
\label{age_sp}
\end{figure*}
Spatial variation in stellar ages with respect to the cluster radius is a footprint of how star formation progressed throughout space and time.
Along with kinematics of the stars, it traces the duration of the star formation process, therefore helps  to constrain 
theoretical models and improve our understanding on the cluster formation.
For example, in ONC, \citet{fur08} and \citet{tob09} showed that
the radial velocities of most of the optically studied stars  are consistent with those of the molecular
gas, which led them to  suggest that both the gas and stars in ONC are collapsing towards the central regions. They argue that  
ONC is kinematically young and yet not dynamically relaxed and  in a state of dynamical collapse. 
Similarly, a decade ago \citet{hart07} suggested that the Orion A cloud 
is undergoing gravitational collapse on large scales, and is producing ONC through the focusing effects of gravity. 
Their suggestion is that the effects of the global gravity result in ever increasing densities with runaway contraction 
in subregions and 
a small number of stars apparently older than a few Myr found in or projected upon 
star-forming regions may be a signature of the cloud evolution.  
The idea of large-scale gravitational collapse is also consistent with observed column 
density probability density functions \citep{bal11}. If the cluster forming clumps are in a state
of hierarchical runaway global gravitational contraction then one would expect older stars at 
the outskirts of the cluster than near the center \citep[e.g., see discussion in][]{sem17}. In fact, based on photometric analyses such a core-halo age 
gradient has been seen in the ONC: the PMS stars in the cluster core appear younger and thus
were formed later than the PMS stars in the cluster halo \citep{get14b}, consistent 
with 
the age segregation noticed in the ONC by \citet{hil97} three decades ago. 

Figure \ref{age_sp} shows the spatial distribution of the stellar ages of the low-mass (0.3 to 1.5 M$_\odot$) stars in Be~59, 
in which the blue to yellow colors represent the age sequence from young to old. 
As can be seen from the figure, the inner region of the cluster shows some degree of segregation of younger stars 
(i.e, redder sources) compared to the outer region. A Two-sample Kolmogorov - Smirnov (KS) test confirms that the age distributions of the stars within 0 to 2 arcmin and 2 to 4 arcmin are distinct with a p-value of 
0.01. Since our age analysis is based only on photometric data, 
a further investigation is certainly needed. In any case, if this `age segregation' is fully confirmed, this may have 
significant implications on the formation of Be~59. 
\begin{figure}
\centering
\includegraphics[scale = 0.3, trim = 5 5 5  5, clip]{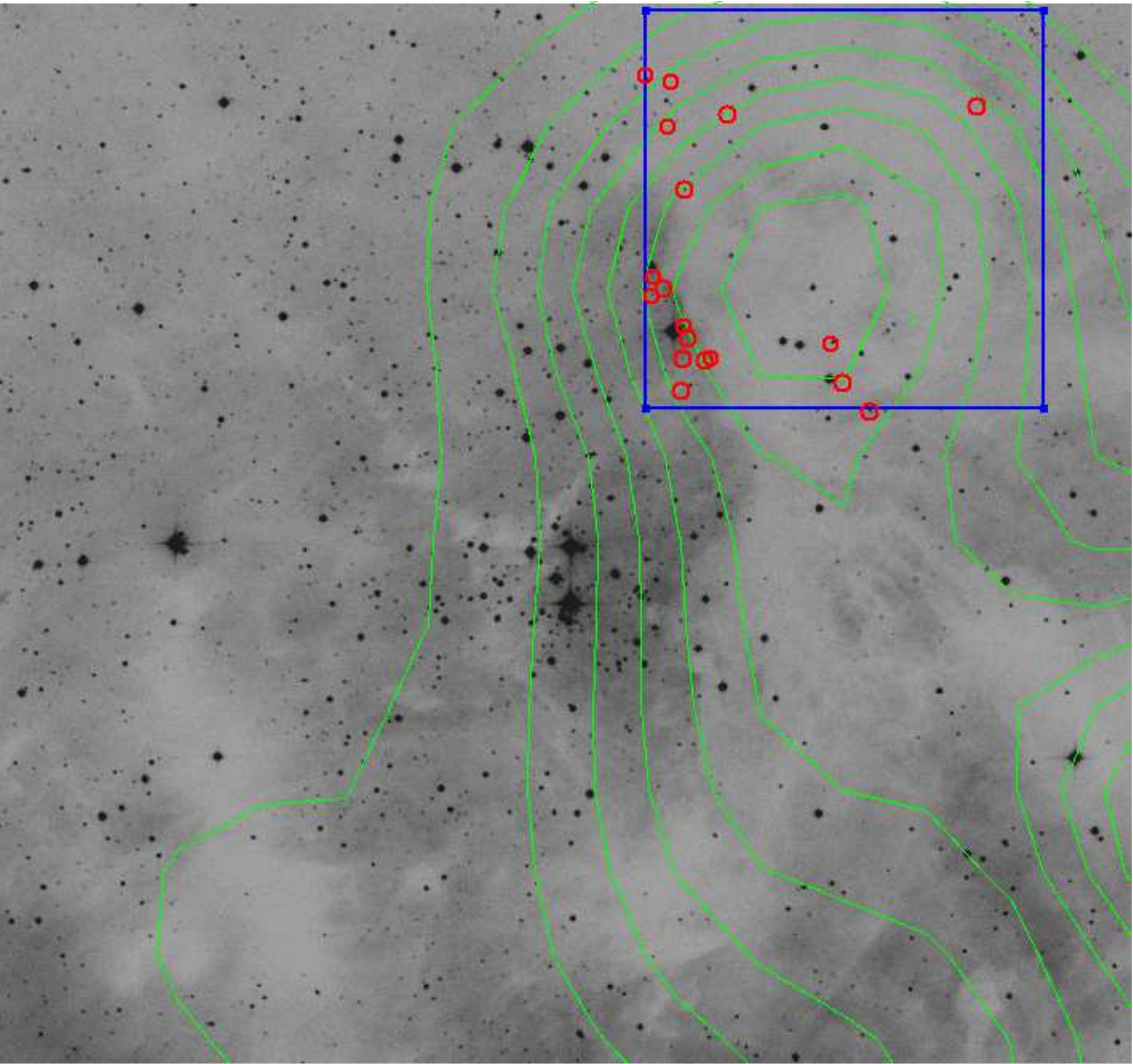}
\includegraphics[scale = 0.34, trim = 5 5 5  5, clip]{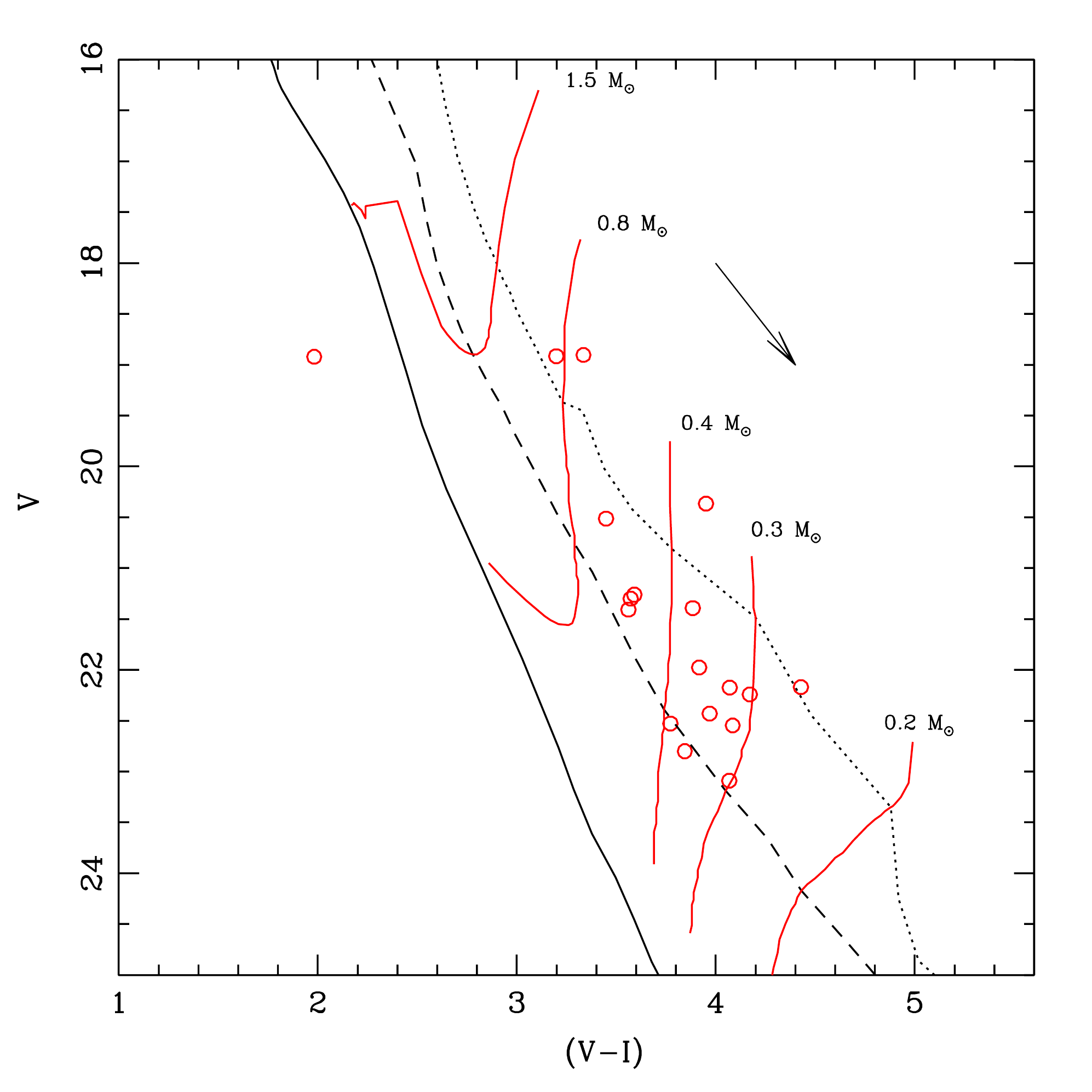}
\caption{Upper panel: The spatial distribution of the YSOs (red circles) in region B along with the cloud dust emission at 
870 GHz observed with $Planck$ (green contours). Lower panel: The locations 
of the YSOs on the $V$ vs. ($V$ - $I$) CMD. The isochrones and evolutionary tracks have the same meaning as in Fig. 4.
}
\label{sf}
\end{figure}
\subsection{Triggered star formation around Be~59}
The Be 59 cluster lies in the center of the H{\sc ii} region Sh2-171, also known as W1. Bright-rimmed clouds BRC~1, BRC~2 and BRC~3 are to the west, 
north and east of the cluster, respectively \citep[cf.][]{sug91}. 
BRCs are thought to arise under the impact of UV photons from nearby massive stars on pre-existing 
dense molecular material \citep{lef94,mia09}, leading to the formation of a new generation of stars by a radiatively-driven implosion \citep{ogu02,mor04,sam12,pan14,sha16}.
\citet{yan92} 
found two dense molecular clouds (`C1' and `C2') on the western side of Be~59 (see Fig. 1).
One of the clumps C1 harbours BRC~1. 
\citet{ogu02} detected six H$\alpha$ stars in the vicinity of BRC~1 may be low-mass stars of the second generation. 
A small elongated aggregate of stars inside the clump C2 has been noticed by \citet{ros13}. In the present work,  we have studied the region B which 
encloses C2. Our optical observations detected counterparts of a few young stars \citep{get17} in region B. The spatial distribution of these 
stars along with the cloud dust emission at 870 GHz is shown in Fig. \ref{sf} (left panel) and their locations on the $V$ vs. ($V$ - $I$) CMD are shown in 
Fig. \ref{sf} (right panel). In Fig. \ref{sf} (right panel), the PMS isochrones of 1 and 5 Myr from \citet{sie00}, corrected for the adopted distance and reddening of the cluster, are also plotted.
 As can be seen from Fig. \ref{sf}, these stars are mostly located towards the south of the clump which
 is close to the cluster. 
We note these stars  are located well within the cluster radius of Be~59, thus, these stars could be part of the cluster. We assessed the age and mass of these YSOs using the same approach as discussed in Sec. 3.3. 
We estimated a mean age of these stars as $\sim$ 2.5 Myr. This mean age is normally higher 
than that of the cluster (i.e., 1.8 Myr). 
We note that adoption of a higher 
extinction value would yield even older ages for these stars. Thus, probably 
being older (or of similar age) than the members of the cluster, they are unlikely 
to be formed as a result of the triggering by the massive 
stars of Be~59. 
Our present data is not sensitive enough to pin-down star formation within the clump C2. 
Deep infra-red photometric and spectroscopic observations 
would shed more light on the star formation and evolutionary stages of the YSOs within C2. 
\section{Conclusions}
We present the $V$ and $I$ band observations of the central region of the cluster Berkeley 59 using the 
3.58-m Telescopio Nazionale Galileo, which are the deepest ($V$ $\sim$ 24 mag) observations so far for the region. 
The optical $V$ vs. ($V$-$I$) CMD for the stars in the region shows a well defined PMS population. 
We estimated the extinction of A$_V$ = 4 mag towards the cluster using the NIR (J - H) and (H - K) colours of the probable PMS 
low-mass stars. We assessed the age and mass of the PMS population identified in the present work by comparing their locations on the CMD 
with the evolutionary models of various ages by 
\citet{sie00}. This yield a mean age of $\sim$ 1.8 Myr 
for the cluster, which is comparable to that of previous studies. 
We find that the slopes of the IMF for the stars in the cluster region are -1.33 and -1.23 in the mass ranges 
0.2 - 28 M$_\odot$ and 0.2 - 1.5  M$_\odot$, respectively, consistent with the other Galactic clusters. 
To study the radial variation of the mass function within the cluster, we constructed the IMFs in two regions: inner 
(0 - 2 arcmin) and outer (2 - 4 arcmin). The slope of the IMF is flatter in the inner region 
compared to that of the outer region, indicating that there are relatively more low-mass stars in the outer region as 
compare to the inner region. Since it is a young cluster ($<$ 2 Myr), still 
associated with the parental cloud, the observed mass-segregation is likely primordial.
The age distribution of the PMS sources reveals that most of the younger sources are centrally distributed 
and older stars tend to be in the outer region of the cluster.
Comparing the mass-radius relation of Be~59 with other young clusters, it seems 
to resemble the 
Trapezium cluster. Our present data do not show signs of triggering and age sequence towards the clump C2. The deep 
infra-red observations of the clump can shed more light on the embedded young stellar population and its properties.  
\acknowledgments
We are thankful to the anonymous referee for valuable comments. 
NP acknowledges the financial support from the Department of Science \& Technology,
INDIA, through INSPIRE faculty award~IFA-PH-36 at University of Delhi. This work is based on observations made with the Italian Telescopio Nazionale Galileo (TNG) operated on the island of La Palma by the Fundación Galileo Galilei of the INAF (Istituto Nazionale di Astrofisica) at the Spanish Observatorio del Roque de los Muchachos of the Instituto de Astrofisica de Canarias. This publication makes use of data from the Two Micron All Sky Survey (a 
joint project of the University of Massachusetts and the Infrared Processing 
and Analysis Center/ California Institute of Technology, funded by the 
National Aeronautics and Space Administration and the National Science 
Foundation), archival data obtained with the {\it Spitzer Space Telescope}
(operated by the Jet Propulsion Laboratory, California Institute 
of Technology, under contract with the NASA.

\bibliography{r}

\begin{thebibliography}{}
\expandafter\ifx\csname natexlab\endcsname\relax\def\natexlab#1{#1}\fi

\bibitem[{{Andersen} {et~al.}(2008){Andersen}, {Meyer}, {Greissl}, \&
  {Aversa}}]{and08}
{Andersen}, M., {Meyer}, M.~R., {Greissl}, J., \& {Aversa}, A. 2008, ApJL, 683,
  L183

\bibitem[{{Andersen} {et~al.}(2011){Andersen}, {Meyer}, {Robberto}, {Bergeron},
  \& {Reid}}]{and11}
{Andersen}, M., {Meyer}, M.~R., {Robberto}, M., {Bergeron}, L.~E., \& {Reid},
  N. 2011, A\&A, 534, A10

\bibitem[{{Andr{\'e}}(1995)}]{and95}
{Andr{\'e}}, P. 1995, Astrophysics \& Space Science, 224, 29

\bibitem[{{Ballesteros-Paredes} {et~al.}(2011){Ballesteros-Paredes},
  {V{\'a}zquez-Semadeni}, {Gazol}, {Hartmann}, {Heitsch}, \&
  {Col{\'{\i}}n}}]{bal11}
{Ballesteros-Paredes}, J., {V{\'a}zquez-Semadeni}, E., {Gazol}, A., {et~al.}
  2011, MNRAS, 416, 1436

\bibitem[{{Baraffe} {et~al.}(1998){Baraffe}, {Chabrier}, {Allard}, \&
  {Hauschildt}}]{bar98}
{Baraffe}, I., {Chabrier}, G., {Allard}, F., \& {Hauschildt}, P.~H. 1998, A\&A,
  337, 403

\bibitem[{{Bastian} {et~al.}(2010){Bastian}, {Covey}, \& {Meyer}}]{bast10}
{Bastian}, N., {Covey}, K.~R., \& {Meyer}, M.~R. 2010, ARA\&A, 48, 339

\bibitem[{{Bate}(2009)}]{bate09}
{Bate}, M.~R. 2009, MNRAS, 392, 1363

\bibitem[{{Bessell} \& {Brett}(1988)}]{bes88}
{Bessell}, M.~S., \& {Brett}, J.~M. 1988, PASP, 100, 1134

\bibitem[{{Bonnell} \& {Bate}(2006)}]{bonn06}
{Bonnell}, I.~A., \& {Bate}, M.~R. 2006, MNRAS, 370, 488

\bibitem[{{Bonnell} {et~al.}(2001){Bonnell}, {Clarke}, {Bate}, \&
  {Pringle}}]{bonn01}
{Bonnell}, I.~A., {Clarke}, C.~J., {Bate}, M.~R., \& {Pringle}, J.~E. 2001,
  MNRAS, 324, 573

\bibitem[{{Bonnell} \& {Davies}(1998)}]{bonn98}
{Bonnell}, I.~A., \& {Davies}, M.~B. 1998, MNRAS, 295, 691

\bibitem[{{Brandner} {et~al.}(2008){Brandner}, {Clark}, {Stolte}, {Waters},
  {Negueruela}, \& {Goodwin}}]{bran08}
{Brandner}, W., {Clark}, J.~S., {Stolte}, A., {et~al.} 2008, A\&A, 478, 137

\bibitem[{{Brice{\~n}o} {et~al.}(2002){Brice{\~n}o}, {Luhman}, {Hartmann},
  {Stauffer}, \& {Kirkpatrick}}]{bri02}
{Brice{\~n}o}, C., {Luhman}, K.~L., {Hartmann}, L., {Stauffer}, J.~R., \&
  {Kirkpatrick}, J.~D. 2002, ApJ, 580, 317

\bibitem[{{Broos} {et~al.}(2013){Broos}, {Getman}, {Povich}, {Feigelson},
  {Townsley}, {Naylor}, {Kuhn}, {King}, \& {Busk}}]{bro13}
{Broos}, P.~S., {Getman}, K.~V., {Povich}, M.~S., {et~al.} 2013, ApJS, 209, 32

\bibitem[{{Burningham} {et~al.}(2005){Burningham}, {Naylor}, {Littlefair}, \&
  {Jeffries}}]{bur05}
{Burningham}, B., {Naylor}, T., {Littlefair}, S.~P., \& {Jeffries}, R.~D. 2005,
  MNRAS, 363, 1389

\bibitem[{{Carpenter}(2001)}]{car01}
{Carpenter}, J.~M. 2001, AJ, 121, 2851

\bibitem[{{Chauhan} {et~al.}(2011){Chauhan}, {Pandey}, {Ogura}, {Jose}, {Ojha},
  {Samal}, \& {Mito}}]{cha11}
{Chauhan}, N., {Pandey}, A.~K., {Ogura}, K., {et~al.} 2011, MNRAS, 415, 1202

\bibitem[{{Chauhan} {et~al.}(2009){Chauhan}, {Pandey}, {Ogura}, {Ojha},
  {Bhatt}, {Ghosh}, \& {Rawat}}]{cha09}
---. 2009, MNRAS, 396, 964

\bibitem[{{Chen} {et~al.}(2007){Chen}, {de Grijs}, \& {Zhao}}]{chen07}
{Chen}, L., {de Grijs}, R., \& {Zhao}, J.~L. 2007, AJ, 134, 1368

\bibitem[{{Cohen} {et~al.}(1981){Cohen}, {Persson}, {Elias}, \&
  {Frogel}}]{coh81}
{Cohen}, J.~G., {Persson}, S.~E., {Elias}, J.~H., \& {Frogel}, J.~A. 1981, ApJ,
  249, 481

\bibitem[{{Da Rio} {et~al.}(2010){Da Rio}, {Robberto}, {Soderblom}, {Panagia},
  {Hillenbrand}, {Palla}, \& {Stassun}}]{da10}
{Da Rio}, N., {Robberto}, M., {Soderblom}, D.~R., {et~al.} 2010, ApJ, 722, 1092

\bibitem[{{Deharveng} {et~al.}(2012){Deharveng}, {Zavagno}, {Anderson},
  {Motte}, {Abergel}, {Andr{\'e}}, {Bontemps}, {Leleu}, {Roussel}, \&
  {Russeil}}]{deh12}
{Deharveng}, L., {Zavagno}, A., {Anderson}, L.~D., {et~al.} 2012, A\&A, 546,
  A74

\bibitem[{{Dib} {et~al.}(2010){Dib}, {Shadmehri}, {Padoan}, {Maheswar}, {Ojha},
  \& {Khajenabi}}]{dib10}
{Dib}, S., {Shadmehri}, M., {Padoan}, P., {et~al.} 2010, MNRAS, 405, 401

\bibitem[{{Elmegreen}(2000)}]{elm00}
{Elmegreen}, B.~G. 2000, ApJ, 530, 277

\bibitem[{{Elmegreen}(2007)}]{elm07}
---. 2007, ApJ, 668, 1064

\bibitem[{{Eswaraiah} {et~al.}(2012){Eswaraiah}, {Pandey}, {Maheswar}, {Chen},
  {Ojha}, \& {Chandola}}]{esw12}
{Eswaraiah}, C., {Pandey}, A.~K., {Maheswar}, G., {et~al.} 2012, MNRAS, 419,
  2587

\bibitem[{{Feigelson} \& {Montmerle}(1999)}]{fei99}
{Feigelson}, E.~D., \& {Montmerle}, T. 1999, ARA\&A, 37, 363

\bibitem[{{Feigelson} {et~al.}(2013){Feigelson}, {Townsley}, {Broos}, {Busk},
  {Getman}, {King}, {Kuhn}, {Naylor}, {Povich}, {Baddeley}, {Bate},
  {Indebetouw}, {Luhman}, {McCaughrean}, {Pittard}, {Pudritz}, {Sills}, {Song},
  \& {Wadsley}}]{fei13}
{Feigelson}, E.~D., {Townsley}, L.~K., {Broos}, P.~S., {et~al.} 2013, ApJS,
  209, 26

\bibitem[{{F{\H u}r{\'e}sz} {et~al.}(2008){F{\H u}r{\'e}sz}, {Hartmann},
  {Megeath}, {Szentgyorgyi}, \& {Hamden}}]{fur08}
{F{\H u}r{\'e}sz}, G., {Hartmann}, L.~W., {Megeath}, S.~T., {Szentgyorgyi},
  A.~H., \& {Hamden}, E.~T. 2008, ApJ, 676, 1109

\bibitem[{{Gennaro} {et~al.}(2011){Gennaro}, {Brandner}, {Stolte}, \&
  {Henning}}]{gen11}
{Gennaro}, M., {Brandner}, W., {Stolte}, A., \& {Henning}, T. 2011, MNRAS, 412,
  2469

\bibitem[{{Getman} {et~al.}(2017){Getman}, {Broos}, {Kuhn}, {Feigelson},
  {Richert}, {Ota}, {Bate}, \& {Garmire}}]{get17}
{Getman}, K.~V., {Broos}, P.~S., {Kuhn}, M.~A., {et~al.} 2017, ApJS, 229, 28

\bibitem[{{Getman} {et~al.}(2005){Getman}, {Feigelson}, {Grosso},
  {McCaughrean}, {Micela}, {Broos}, {Garmire}, \& {Townsley}}]{get05}
{Getman}, K.~V., {Feigelson}, E.~D., {Grosso}, N., {et~al.} 2005, ApJS, 160,
  353

\bibitem[{{Getman} {et~al.}(2014){Getman}, {Feigelson}, \& {Kuhn}}]{get14b}
{Getman}, K.~V., {Feigelson}, E.~D., \& {Kuhn}, M.~A. 2014, ApJ, 787, 109

\bibitem[{{Getman} {et~al.}(2012){Getman}, {Feigelson}, {Sicilia-Aguilar},
  {Broos}, {Kuhn}, \& {Garmire}}]{getman12}
{Getman}, K.~V., {Feigelson}, E.~D., {Sicilia-Aguilar}, A., {et~al.} 2012,
  MNRAS, 426, 2917

\bibitem[{{Girardi} {et~al.}(2002){Girardi}, {Bertelli}, {Bressan}, {Chiosi},
  {Groenewegen}, {Marigo}, {Salasnich}, \& {Weiss}}]{gir02}
{Girardi}, L., {Bertelli}, G., {Bressan}, A., {et~al.} 2002, A\&A, 391, 195

\bibitem[{{G{\"u}del} {et~al.}(2007){G{\"u}del}, {Skinner}, {Mel'Nikov},
  {Audard}, {Telleschi}, \& {Briggs}}]{gue07}
{G{\"u}del}, M., {Skinner}, S.~L., {Mel'Nikov}, S.~Y., {et~al.} 2007, A\&A,
  468, 529

\bibitem[{{Hartmann} \& {Burkert}(2007{\natexlab{a}})}]{har07}
{Hartmann}, L., \& {Burkert}, A. 2007{\natexlab{a}}, ApJ, 654, 988

\bibitem[{{Hartmann} \& {Burkert}(2007{\natexlab{b}})}]{hart07}
---. 2007{\natexlab{b}}, ApJ, 654, 988

\bibitem[{{Hayes} {et~al.}(2015){Hayes}, {Friel}, {Slack}, \& {Boberg}}]{hay15}
{Hayes}, C.~R., {Friel}, E.~D., {Slack}, T.~J., \& {Boberg}, O.~M. 2015, AJ,
  150, 200

\bibitem[{{Herbst} {et~al.}(1994){Herbst}, {Herbst}, {Grossman}, \&
  {Weinstein}}]{her94}
{Herbst}, W., {Herbst}, D.~K., {Grossman}, E.~J., \& {Weinstein}, D. 1994, AJ,
  108, 1906

\bibitem[{{Hillenbrand}(1997)}]{hil97}
{Hillenbrand}, L.~A. 1997, AJ, 113, 1733

\bibitem[{{Hillenbrand} \& {Carpenter}(2000)}]{hill00}
{Hillenbrand}, L.~A., \& {Carpenter}, J.~M. 2000, ApJ, 540, 236

\bibitem[{{Jeffries} {et~al.}(2011){Jeffries}, {Littlefair}, {Naylor}, \&
  {Mayne}}]{jeff11}
{Jeffries}, R.~D., {Littlefair}, S.~P., {Naylor}, T., \& {Mayne}, N.~J. 2011,
  MNRAS, 418, 1948

\bibitem[{{Jose} {et~al.}(2017){Jose}, {Herczeg}, {Samal}, {Fang}, \&
  {Panwar}}]{jos17}
{Jose}, J., {Herczeg}, G.~J., {Samal}, M.~R., {Fang}, Q., \& {Panwar}, N. 2017,
  ApJ, 836, 98

\bibitem[{{Jose} {et~al.}(2016){Jose}, {Kim}, {Herczeg}, {Samal}, {Bieging},
  {Meyer}, \& {Sherry}}]{jose16}
{Jose}, J., {Kim}, J.~S., {Herczeg}, G.~J., {et~al.} 2016, ApJ, 822, 49

\bibitem[{{Jose} {et~al.}(2008){Jose}, {Pandey}, {Ojha}, {Ogura}, {Chen},
  {Bhatt}, {Ghosh}, {Mito}, {Maheswar}, \& {Sharma}}]{jos08}
{Jose}, J., {Pandey}, A.~K., {Ojha}, D.~K., {et~al.} 2008, MNRAS, 384, 1675

\bibitem[{{Jose} {et~al.}(2013){Jose}, {Pandey}, {Samal}, {Ojha}, {Ogura},
  {Kim}, {Kobayashi}, {Goyal}, {Chauhan}, \& {Eswaraiah}}]{jos13}
{Jose}, J., {Pandey}, A.~K., {Samal}, M.~R., {et~al.} 2013, MNRAS, 432, 3445

\bibitem[{{Koenig} {et~al.}(2008){Koenig}, {Allen}, {Gutermuth}, {Hora},
  {Brunt}, \& {Muzerolle}}]{koe08}
{Koenig}, X.~P., {Allen}, L.~E., {Gutermuth}, R.~A., {et~al.} 2008, ApJ, 688,
  1142

\bibitem[{{Koenig} {et~al.}(2012){Koenig}, {Leisawitz}, {Benford}, {Rebull},
  {Padgett}, \& {Assef}}]{koe12}
{Koenig}, X.~P., {Leisawitz}, D.~T., {Benford}, D.~J., {et~al.} 2012, ApJ, 744,
  130

\bibitem[{{Kroupa} {et~al.}(2001){Kroupa}, {Aarseth}, \& {Hurley}}]{kro01}
{Kroupa}, P., {Aarseth}, S., \& {Hurley}, J. 2001, MNRAS, 321, 699

\bibitem[{{Kun} {et~al.}(2008){Kun}, {Kiss}, \& {Balog}}]{kun08}
{Kun}, M., {Kiss}, Z.~T., \& {Balog}, Z. 2008, {Star Forming Regions in
  Cepheus}, ed. B.~{Reipurth}, 136

\bibitem[{{Lada}(1987)}]{lada87}
{Lada}, C.~J. 1987, in IAU Symposium, Vol. 115, Star Forming Regions, ed.
  M.~{Peimbert} \& J.~{Jugaku}, 1--17

\bibitem[{{Lada} \& {Lada}(2003)}]{lada03}
{Lada}, C.~J., \& {Lada}, E.~A. 2003, ARA\&A, 41, 57

\bibitem[{{Lada} {et~al.}(2006){Lada}, {Muench}, {Luhman}, {Allen}, {Hartmann},
  {Megeath}, {Myers}, {Fazio}, {Wood}, {Muzerolle}, {Rieke}, {Siegler}, \&
  {Young}}]{lada06}
{Lada}, C.~J., {Muench}, A.~A., {Luhman}, K.~L., {et~al.} 2006, AJ, 131, 1574

\bibitem[{{Lata} {et~al.}(2012){Lata}, {Pandey}, {Chen}, {Maheswar}, \&
  {Chauhan}}]{lat12}
{Lata}, S., {Pandey}, A.~K., {Chen}, W.~P., {Maheswar}, G., \& {Chauhan}, N.
  2012, MNRAS, 427, 1449

\bibitem[{{Lata} {et~al.}(2011){Lata}, {Pandey}, {Maheswar}, {Mondal}, \&
  {Kumar}}]{lat11}
{Lata}, S., {Pandey}, A.~K., {Maheswar}, G., {Mondal}, S., \& {Kumar}, B. 2011,
  MNRAS, 418, 1346

\bibitem[{{Lata} {et~al.}(2016){Lata}, {Pandey}, {Panwar}, {Chen}, {Samal}, \&
  {Pandey}}]{lat15}
{Lata}, S., {Pandey}, A.~K., {Panwar}, N., {et~al.} 2016, MNRAS, 456, 2505

\bibitem[{{Lefloch} \& {Lazareff}(1994)}]{lef94}
{Lefloch}, B., \& {Lazareff}, B. 1994, A\&A, 289, 559

\bibitem[{{Lim} {et~al.}(2013){Lim}, {Chun}, {Sung}, {Park}, {Lee}, {Sohn},
  {Hur}, \& {Bessell}}]{lim13}
{Lim}, B., {Chun}, M.-Y., {Sung}, H., {et~al.} 2013, AJ, 145, 46

\bibitem[{{Liu} {et~al.}(1989){Liu}, {Janes}, \& {Bania}}]{liu89}
{Liu}, T., {Janes}, K.~A., \& {Bania}, T.~M. 1989, AJ, 98, 626

\bibitem[{{Liu} {et~al.}(1988){Liu}, {Janes}, {Bania}, \& {Phelps}}]{liu88}
{Liu}, T., {Janes}, K.~A., {Bania}, T.~M., \& {Phelps}, R.~L. 1988, AJ, 95,
  1122

\bibitem[{{Lu} {et~al.}(2013){Lu}, {Do}, {Ghez}, {Morris}, {Yelda}, \&
  {Matthews}}]{lu13}
{Lu}, J.~R., {Do}, T., {Ghez}, A.~M., {et~al.} 2013, ApJ, 764, 155

\bibitem[{{Lucas} {et~al.}(2005){Lucas}, {Roche}, \& {Tamura}}]{luc05}
{Lucas}, P.~W., {Roche}, P.~F., \& {Tamura}, M. 2005, MNRAS, 361, 211

\bibitem[{{Luhman}(2004)}]{luh04}
{Luhman}, K.~L. 2004, ApJ, 617, 1216

\bibitem[{{Luhman}(2007)}]{luh07}
---. 2007, ApJS, 173, 104

\bibitem[{{Luhman}(2012)}]{luh12r}
---. 2012, ARA\&A, 50, 65

\bibitem[{{Luhman} {et~al.}(2003{\natexlab{a}}){Luhman}, {Brice{\~n}o},
  {Stauffer}, {Hartmann}, {Barrado y Navascu{\'e}s}, \& {Caldwell}}]{luh03a}
{Luhman}, K.~L., {Brice{\~n}o}, C., {Stauffer}, J.~R., {et~al.}
  2003{\natexlab{a}}, ApJ, 590, 348

\bibitem[{{Luhman} {et~al.}(2003{\natexlab{b}}){Luhman}, {Stauffer}, {Muench},
  {Rieke}, {Lada}, {Bouvier}, \& {Lada}}]{luh03b}
{Luhman}, K.~L., {Stauffer}, J.~R., {Muench}, A.~A., {et~al.}
  2003{\natexlab{b}}, ApJ, 593, 1093

\bibitem[{{Ma{\'{\i}}z Apell{\'a}niz} {et~al.}(2016){Ma{\'{\i}}z
  Apell{\'a}niz}, {Sota}, {Arias}, {Barb{\'a}}, {Walborn},
  {Sim{\'o}n-D{\'{\i}}az}, {Negueruela}, {Marco}, {Le{\~a}o}, {Herrero},
  {Gamen}, \& {Alfaro}}]{mai16}
{Ma{\'{\i}}z Apell{\'a}niz}, J., {Sota}, A., {Arias}, J.~I., {et~al.} 2016,
  ApJS, 224, 4

\bibitem[{{Majaess} {et~al.}(2008){Majaess}, {Turner}, {Lane}, \&
  {Moncrieff}}]{maj08}
{Majaess}, D.~J., {Turner}, D.~G., {Lane}, D.~J., \& {Moncrieff}, K.~E. 2008,
  Journal of the American Association of Variable Star Observers (JAAVSO), 36,
  90

\bibitem[{{McMillan} {et~al.}(2007){McMillan}, {Vesperini}, \& {Portegies
  Zwart}}]{mcm07}
{McMillan}, S.~L.~W., {Vesperini}, E., \& {Portegies Zwart}, S.~F. 2007, ApJL,
  655, L45

\bibitem[{{Meyer} {et~al.}(1997){Meyer}, {Calvet}, \& {Hillenbrand}}]{mey97a}
{Meyer}, M.~R., {Calvet}, N., \& {Hillenbrand}, L.~A. 1997, AJ, 114, 288

\bibitem[{{Miao} {et~al.}(2009){Miao}, {White}, {Thompson}, \&
  {Nelson}}]{mia09}
{Miao}, J., {White}, G.~J., {Thompson}, M.~A., \& {Nelson}, R.~P. 2009, ApJ,
  692, 382

\bibitem[{{Morgan} {et~al.}(2004){Morgan}, {Thompson}, {Urquhart}, {White}, \&
  {Miao}}]{mor04}
{Morgan}, L.~K., {Thompson}, M.~A., {Urquhart}, J.~S., {White}, G.~J., \&
  {Miao}, J. 2004, A\&A, 426, 535

\bibitem[{{Muench} {et~al.}(2002){Muench}, {Lada}, {Lada}, \& {Alves}}]{mue02}
{Muench}, A.~A., {Lada}, E.~A., {Lada}, C.~J., \& {Alves}, J. 2002, ApJ, 573,
  366

\bibitem[{{Muench} {et~al.}(2003){Muench}, {Lada}, {Lada}, {Elston}, {Alves},
  {Horrobin}, {Huard}, {Levine}, {Raines}, \&
  {Rom{\'a}n-Z{\'u}{\~n}iga}}]{mue03}
{Muench}, A.~A., {Lada}, E.~A., {Lada}, C.~J., {et~al.} 2003, AJ, 125, 2029

\bibitem[{{Neichel} {et~al.}(2015){Neichel}, {Samal}, {Plana}, {Zavagno},
  {Bernard}, \& {Fusco}}]{nei15}
{Neichel}, B., {Samal}, M.~R., {Plana}, H., {et~al.} 2015, A\&A, 576, A110

\bibitem[{{Offner} {et~al.}(2014){Offner}, {Clark}, {Hennebelle}, {Bastian},
  {Bate}, {Hopkins}, {Moraux}, \& {Whitworth}}]{off14}
{Offner}, S.~S.~R., {Clark}, P.~C., {Hennebelle}, P., {et~al.} 2014, Protostars
  and Planets VI, 53

\bibitem[{{Ogura} {et~al.}(2002){Ogura}, {Sugitani}, \& {Pickles}}]{ogu02}
{Ogura}, K., {Sugitani}, K., \& {Pickles}, A. 2002, AJ, 123, 2597

\bibitem[{{Ojha} {et~al.}(2009){Ojha}, {Tamura}, {Nakajima}, {Saito}, {Pandey},
  {Ghosh}, \& {Aoki}}]{ojha09}
{Ojha}, D.~K., {Tamura}, M., {Nakajima}, Y., {et~al.} 2009, ApJ, 693, 634

\bibitem[{{Ojha} {et~al.}(2011){Ojha}, {Samal}, {Pandey}, {Bhatt}, {Ghosh},
  {Sharma}, {Tamura}, {Mohan}, \& {Zinchenko}}]{ojha11}
{Ojha}, D.~K., {Samal}, M.~R., {Pandey}, A.~K., {et~al.} 2011, ApJ, 738, 156

\bibitem[{{Oliveira} {et~al.}(2009){Oliveira}, {Jeffries}, \& {van
  Loon}}]{oli09}
{Oliveira}, J.~M., {Jeffries}, R.~D., \& {van Loon}, J.~T. 2009, MNRAS, 392,
  1034

\bibitem[{{Pandey} {et~al.}(2001){Pandey}, {Nilakshi}, {Ogura}, {Sagar}, \&
  {Tarusawa}}]{pan01}
{Pandey}, A.~K., {Nilakshi}, {Ogura}, K., {Sagar}, R., \& {Tarusawa}, K. 2001,
  A\&A, 374, 504

\bibitem[{{Pandey} {et~al.}(2013{\natexlab{a}}){Pandey}, {Samal}, {Chauhan},
  {Eswaraiah}, {Pandey}, {Chen}, \& {Ojha}}]{pan13}
{Pandey}, A.~K., {Samal}, M.~R., {Chauhan}, N., {et~al.} 2013{\natexlab{a}},
  New Astronomy, 19, 1

\bibitem[{{Pandey} {et~al.}(2014){Pandey}, {Samal}, {Yadav}, {Richichi},
  {Lata}, {Pandey}, {Ojha}, \& {Chen}}]{pand14}
{Pandey}, A.~K., {Samal}, M.~R., {Yadav}, R.~K., {et~al.} 2014, New Astronomy,
  29, 18

\bibitem[{{Pandey} {et~al.}(2008){Pandey}, {Sharma}, {Ogura}, {Ojha}, {Chen},
  {Bhatt}, \& {Ghosh}}]{pan08}
{Pandey}, A.~K., {Sharma}, S., {Ogura}, K., {et~al.} 2008, MNRAS, 383, 1241

\bibitem[{{Pandey} {et~al.}(2013{\natexlab{b}}){Pandey}, {Eswaraiah}, {Sharma},
  {Samal}, {Chauhan}, {Chen}, {Jose}, {Ojha}, {Kesh Yadav}, \&
  {Chandola}}]{pan13a}
{Pandey}, A.~K., {Eswaraiah}, C., {Sharma}, S., {et~al.} 2013{\natexlab{b}},
  ApJ, 764, 172

\bibitem[{{Pang} {et~al.}(2013){Pang}, {Grebel}, {Allison}, {Goodwin},
  {Altmann}, {Harbeck}, {Moffat}, \& {Drissen}}]{pang13}
{Pang}, X., {Grebel}, E.~K., {Allison}, R.~J., {et~al.} 2013, ApJ, 764, 73

\bibitem[{{Panwar} {et~al.}(2014){Panwar}, {Chen}, {Pandey}, {Samal}, {Ogura},
  {Ojha}, {Jose}, \& {Bhatt}}]{pan14}
{Panwar}, N., {Chen}, W.~P., {Pandey}, A.~K., {et~al.} 2014, MNRAS, 443, 1614

\bibitem[{{Panwar} {et~al.}(2017){Panwar}, {Samal}, {Pandey}, {Jose}, {Chen},
  {Ojha}, {Ogura}, {Singh}, \& {Yadav}}]{pan17}
{Panwar}, N., {Samal}, M.~R., {Pandey}, A.~K., {et~al.} 2017, MNRAS, 468, 2684

\bibitem[{{Pfalzner}(2011)}]{pfl11}
{Pfalzner}, S. 2011, A\&A, 536, A90

\bibitem[{{Preibisch} \& {Feigelson}(2005)}]{pre05}
{Preibisch}, T., \& {Feigelson}, E.~D. 2005, ApJS, 160, 390

\bibitem[{{Preibisch} {et~al.}(2017){Preibisch}, {Flaischlen}, {Gaczkowski},
  {Townsley}, \& {Broos}}]{pre17}
{Preibisch}, T., {Flaischlen}, S., {Gaczkowski}, B., {Townsley}, L., \&
  {Broos}, P. 2017, A\&A, 605, A85

\bibitem[{{Reggiani} {et~al.}(2011){Reggiani}, {Robberto}, {Da Rio}, {Meyer},
  {Soderblom}, \& {Ricci}}]{regg11}
{Reggiani}, M., {Robberto}, M., {Da Rio}, N., {et~al.} 2011, A\&A, 534, A83

\bibitem[{{Rodr{\'{\i}}guez-Ledesma} {et~al.}(2010){Rodr{\'{\i}}guez-Ledesma},
  {Mundt}, \& {Eisl{\"o}ffel}}]{rod10}
{Rodr{\'{\i}}guez-Ledesma}, M.~V., {Mundt}, R., \& {Eisl{\"o}ffel}, J. 2010,
  A\&A, 515, A13

\bibitem[{{Rosvick} \& {Majaess}(2013)}]{ros13}
{Rosvick}, J.~M., \& {Majaess}, D. 2013, AJ, 146, 142

\bibitem[{{Salpeter}(1955)}]{sal55}
{Salpeter}, E.~E. 1955, ApJ, 121, 161

\bibitem[{{Samal} {et~al.}(2012){Samal}, {Pandey}, {Ojha}, {Chauhan}, {Jose},
  \& {Pandey}}]{sam12}
{Samal}, M.~R., {Pandey}, A.~K., {Ojha}, D.~K., {et~al.} 2012, ApJ, 755, 20

\bibitem[{{Samal} {et~al.}(2014){Samal}, {Zavagno}, {Deharveng}, {Molinari},
  {Ojha}, {Paradis}, {Tig{\'e}}, {Pandey}, \& {Russeil}}]{sam14}
{Samal}, M.~R., {Zavagno}, A., {Deharveng}, L., {et~al.} 2014, A\&A, 566, A122

\bibitem[{{Scholz}(2012)}]{sch12}
{Scholz}, A. 2012, MNRAS, 420, 1495

\bibitem[{{Sharma} {et~al.}(2008){Sharma}, {Pandey}, {Ogura}, {Aoki}, {Pandey},
  {Sandhu}, \& {Sagar}}]{shar08}
{Sharma}, S., {Pandey}, A.~K., {Ogura}, K., {et~al.} 2008, AJ, 135, 1934

\bibitem[{{Sharma} {et~al.}(2016){Sharma}, {Pandey}, {Borissova}, {Ojha},
  {Ivanov}, {Ogura}, {Kobayashi}, {Kurtev}, {Gopinathan}, \& {Kesh
  Yadav}}]{sha16}
{Sharma}, S., {Pandey}, A.~K., {Borissova}, J., {et~al.} 2016, AJ, 151, 126

\bibitem[{{Sicilia-Aguilar} {et~al.}(2005){Sicilia-Aguilar}, {Hartmann},
  {Hern{\'a}ndez}, {Brice{\~n}o}, \& {Calvet}}]{sic05}
{Sicilia-Aguilar}, A., {Hartmann}, L.~W., {Hern{\'a}ndez}, J., {Brice{\~n}o},
  C., \& {Calvet}, N. 2005, AJ, 130, 188

\bibitem[{{Siess} {et~al.}(2000){Siess}, {Dufour}, \& {Forestini}}]{sie00}
{Siess}, L., {Dufour}, E., \& {Forestini}, M. 2000, A\&A, 358, 593

\bibitem[{{Skiff}(2014)}]{skif14}
{Skiff}, B.~A. 2014, VizieR Online Data Catalog, 1

\bibitem[{{Slesnick} {et~al.}(2004){Slesnick}, {Hillenbrand}, \&
  {Carpenter}}]{sle04}
{Slesnick}, C.~L., {Hillenbrand}, L.~A., \& {Carpenter}, J.~M. 2004, ApJ, 610,
  1045

\bibitem[{{Stahl} {et~al.}(2008){Stahl}, {Wade}, {Petit}, {Stober}, \&
  {Schanne}}]{stah08}
{Stahl}, O., {Wade}, G., {Petit}, V., {Stober}, B., \& {Schanne}, L. 2008,
  A\&A, 487, 323

\bibitem[{{Sterzik} \& {Durisen}(2003)}]{ste03}
{Sterzik}, M.~F., \& {Durisen}, R.~H. 2003, A\&A, 400, 1031

\bibitem[{{Stetson}(1987)}]{ste87}
{Stetson}, P.~B. 1987, PASP, 99, 191

\bibitem[{{Stolte} {et~al.}(2005){Stolte}, {Brandner}, {Grebel}, {Lenzen}, \&
  {Lagrange}}]{sto05}
{Stolte}, A., {Brandner}, W., {Grebel}, E.~K., {Lenzen}, R., \& {Lagrange},
  A.-M. 2005, ApJL, 628, L113

\bibitem[{{Sugitani} {et~al.}(1991){Sugitani}, {Fukui}, \& {Ogura}}]{sug91}
{Sugitani}, K., {Fukui}, Y., \& {Ogura}, K. 1991, ApJS, 77, 59

\bibitem[{{Sung} \& {Bessell}(2010)}]{sung10}
{Sung}, H., \& {Bessell}, M.~S. 2010, AJ, 140, 2070

\bibitem[{{Sung} {et~al.}(2013){Sung}, {Sana}, \& {Bessell}}]{sung13}
{Sung}, H., {Sana}, H., \& {Bessell}, M.~S. 2013, AJ, 145, 37

\bibitem[{{Tan} {et~al.}(2006){Tan}, {Krumholz}, \& {McKee}}]{tan06}
{Tan}, J.~C., {Krumholz}, M.~R., \& {McKee}, C.~F. 2006, ApJL, 641, L121

\bibitem[{{Tobin} {et~al.}(2009){Tobin}, {Hartmann}, {Furesz}, {Mateo}, \&
  {Megeath}}]{tob09}
{Tobin}, J.~J., {Hartmann}, L., {Furesz}, G., {Mateo}, M., \& {Megeath}, S.~T.
  2009, ApJ, 697, 1103

\bibitem[{{van Dokkum} \& {Conroy}(2010)}]{van10}
{van Dokkum}, P.~G., \& {Conroy}, C. 2010, Nature, 468, 940

\bibitem[{{V{\'a}zquez-Semadeni} {et~al.}(2017){V{\'a}zquez-Semadeni},
  {Gonz{\'a}lez-Samaniego}, \& {Col{\'{\i}}n}}]{sem17}
{V{\'a}zquez-Semadeni}, E., {Gonz{\'a}lez-Samaniego}, A., \& {Col{\'{\i}}n}, P.
  2017, MNRAS, 467, 1313

\bibitem[{{Wang} {et~al.}(2014){Wang}, {Chen}, {Lin}, {Pandey}, {Huang},
  {Panwar}, {Lee}, {Tsai}, {Tang}, {Goldman}, {Burgett}, {Chambers}, {Draper},
  {Flewelling}, {Grav}, {Heasley}, {Hodapp}, {Huber}, {Jedicke}, {Kaiser},
  {Kudritzki}, {Luppino}, {Lupton}, {Magnier}, {Metcalfe}, {Monet}, {Morgan},
  {Onaka}, {Price}, {Stubbs}, {Sweeney}, {Tonry}, {Wainscoat}, \&
  {Waters}}]{wang14}
{Wang}, P.~F., {Chen}, W.~P., {Lin}, C.~C., {et~al.} 2014, ApJ, 784, 57

\bibitem[{{Yang} \& {Fukui}(1992)}]{yan92}
{Yang}, J., \& {Fukui}, Y. 1992, ApJ, 386, 618

\end{thebibliography}


\listofchanges

\end{document}